\documentclass[11pt,citesort]{article}
\newcommand\ignore[1]{}
\usepackage{amsmath,amssymb,graphicx,url}
\usepackage{color}

\newcommand\be{\begin{equation}}
\newcommand\ee{\end{equation}}
\newcommand\bea{\begin{eqnarray}}
\newcommand\eea{\end{eqnarray}}
\newcommand{\bdm}{\begin{displaymath}}
\newcommand{\edm}{\end{displaymath}}
\newcommand\nn{ \nonumber\\}

\newcommand{\dd}[1]{\partial_{#1}}
\def\dd{\partial}
\newcommand{\<}{\langle}
\renewcommand{\>}{\rangle}

\makeatletter
\@addtoreset{equation}{section}
\makeatother
\renewcommand{\theequation}{\thesection.\arabic{equation}}

\title{Diffractive Higgs Production by AdS Pomeron Fusion}

\author{ Richard  C. Brower~\footnote{Physics Department,
Boston University, Boston MA 02215},
Marko Djuri\'c~\footnote{Centro de F\'\i sica do Porto,
Departamento de F\'\i sica e Astronomia,
Faculdade de Ci\^encias da Universidade do Porto,
4169--007 Porto, Portugal},
and   Chung-I Tan~\footnote{Physics Department, Brown University,
Providence, RI 02912}
}

\begin{document}

\maketitle

\begin{abstract}
  The double diffractive Higgs production at central rapidity is
  formulated in terms of the fusion of two AdS gravitons/Pomerons
first introduced by  Brower, Polchinski, Strassler and Tan in elastic
  scattering.
Here we propose a simple self-consistent holographic framework capable of providing
  phenomenologically compelling estimates of diffractive cross
  sections at the LHC.  As in the traditional weak coupling approach, we
  anticipate that several phenomenological
  parameters  must be tested and calibrated through
  factorization for a self-consistent description of other diffractive
  process such as total cross sections, deep inelastic scattering and
  heavy quark production in the central region.
 \end{abstract}

\newpage
\setcounter{tocdepth}{3}
\tableofcontents

\newpage
\section{Introduction}
\label{sec:intro}

A promising method for studying the Higgs meson at the
LHC involves exclusive double diffractive Higgs production
in forward proton-proton scattering. The protons scatter through very
small angles with large rapidity gaps separating the Higgs in the
central region,
\begin{equation}
 p(k_1) + p(k_2) \rightarrow p(k_3)+  H(q) + p(k_4) \; .
\end{equation}
The Higgs subsequently decays into large transverse momentum
fragments. Although this represents a small fraction of the total
cross section, the exclusive channel should provide an exceptional
signal to background discrimination by constraining the Higgs mass  both to the energy of
decay fragments and to the energy lost to the forward protons~\cite{Kharzeev:2000jwa}. Relaxing the kinematics to allow
for inclusive double diffraction may also be useful, where one or both
of the nucleons are diffractively excited; we will defer these extensions to future studies. While double diffraction is
very unlikely to be a discovery channel, it may play a useful role in
determining properties of the Higgs when and if it is found.

Current phenomenological estimates of the diffractive Higgs production
cross section have generally followed two approaches:  perturbative (weak coupling) vs  confining (strong coupling), or equivalently, in the Regge literature,  often referred to as the ``hard Pomeron''  vs ``soft Pomeron''  methods. Previous works on diffractive Higgs production include \cite{Kharzeev:2000jwa,Khoze:1997dr,Khoze:2000cy,GayDucati:2011zx,Gastmans:2011zz,Bialas1991540,Coughlin:2009tr,Spira:1995rr,Brodsky:2006wb,Ryskin:2009tk} (see for example \cite{GayDucati:2011zx} for additional related references).  The Regge approach to high energy scattering, although well motivated phenomenologically, has suffered in the past by the lack of a precise theoretical underpinning. The advent of AdS/CFT has changed the situation. In a holographic approach,  the Pomeron is a well-defined theoretical
concept. The bare Pomeron is the
leading planar term in the $1/N_c$ expansion at fixed 't Hooft coupling
($\lambda \equiv g^2N_c$) which is  then identified as the ``AdS graviton" in the strong coupling~\cite{Brower:2006ea} limit ($\lambda \rightarrow \infty$).

In this paper, we apply String/Gauge Duality to double diffractive Higgs production  at central rapidity, formulated in terms of the fusion of two  gravitons/Pomerons, first introduced by Brower, Polchinski, Strassler and Tan (BPST)  in~\cite{Brower:2006ea}. High energy diffractive collisions
 already have a rather extensive AdS/CFT  literature  to draw on.  A key observation is  {\bf $\bf AdS$-transverse factorization} that
emerges at high energy as a {\it universal} feature, applicable to scattering involving both particles and currents. This leads to an AdS/Reggeon formulation with a few crucial phenomenological parameters
which need to be fixed experimentally. Consequently using the $AdS$ factorization and  by comparing different processes the parameters are overconstrained allowing
one  both to test the accuracy of the framework and to give confidence to prediction when extended to new  cross sections such as
diffractive Higgs production.   A seminal paper by Strassler and Polchinski on deep inelastic
scattering~\cite{Polchinski:2002jw} already introduced some of the 
results later elaborated in Ref.~\cite{Brower:2006ea} for elastic
scattering.   For instance, for elastic scattering, the amplitude can be represented schematically in a factorizable form (see Eq. (\ref{eq:adsPomeron})), 
\begin{equation}
A(s,t)  = \Phi_{13}(t)*\widetilde {\cal K}_P(s,t) * \Phi_{24}(t) \; ,
\label{eq:adsPomeronScheme}
\end{equation}
where the impact factors $ \Phi_{13}$ and $ \Phi_{24}$ represent two elastic vertex couplings to the external
particles, and $\widetilde {\cal K}_P$ is an universal BPST  Pomeron kernel~\footnote{Unlike the case of a graviton exchange in $AdS$, this Pomeron kernel contains both real and imaginary parts.}, with a characteristic power behavior at large $s>>|t|$, 
\begin{equation}
\widetilde {\cal K}_P\sim s^{j_0}\, .
\label{eq:Regge}
\end{equation} 
This ``Pomeron intercept", $j_0$, lies in the range  $1<j_0<2$ and is a function of the 't Hooft coupling, $g^2N_c$. The $*$-operator is defined  explicitly
by  Eq.~(\ref{eq:adsPomeron}) in Sec.~\ref{sec:Ingredient}  below, with the kernel expressed more explicitly as $\widetilde {\cal K}_P(s,t,z,z')$. It represents a
convolution in the radial coordinate in AdS or more generally, in geometric terms, a 3-d convolution in transverse  space, $({\bf x_\perp},z)$, combining the conventional impact parameter ${\bf x}_\perp$, conjugate to ${\bf k_\perp}$,  and a 3rd radial coordinate, $r \sim 1/z$ of $AdS^5$.
 There is also an extensive literature on eikonal sum in AdS
space~\cite{Brower:2007qh,Brower:2007xg,Cornalba:2006xm,Cornalba:2006xk,Cornalba:2007fs,Cornalba:2007zb}. In a recent paper by Brower, Djuri\'c, Sar\v{c}evi\'c and Tan,  this approach is applied to give a reasonable account of
the small-$x$ contribution to deep inelastic scattering~\cite{Brower:2010wf}.  Here one makes use of the universality property, Eq. (\ref{eq:adsPomeronScheme}). In moving from elastic to DIS, one simply replaces $\Phi_{13}$ in   (\ref{eq:adsPomeronScheme}) by appropriate product of propagators  for external currents~\cite{Polchinski:2002jw,Brower:2010wf}. The same formalism can be applied as well to deeply virtual Compton scattering at small-$x$, as done recently by Costa and Djuri\'c \cite{Costa:2012fw}~\footnote{The factorized form of the amplitude in $AdS$, but without the explicit form of the Pomeron kernel, has also  been  applied  to a subset of DIS data  in \cite{Cornalba:2008sp,Cornalba:2010vk}. }.

By virtue of factorization in AdS space, the extension to double diffractive Higgs production
amplitude takes the form of 
\begin{equation}
A(s,s_1, s_2, t_1, t_2)  = \Phi_{13}(t_1)*  \widetilde{\cal K}_P(t_1,s_1)*V_H(s_1 s_2/s,t_1,t_2)*  \widetilde {\cal K}_P(s_2,t_2)* \Phi_{24}(t_2) \; .
\label{eq:adsDoublePomeronScheme}
\end{equation}
where we introduce the  vertex $V_H$ for the Pomeron fusion to Higgs processes. (See  Eq. (\ref{eq:adsDoublePomeron}) for the explicit form.)  Thus
diffractive Higgs production requires three building blocks: Two from elastic scattering, the proton impact factors, $\Phi_{ij}$ and  the Pomeron kernel (or Reggeon propagators), $\widetilde{\cal K}_P$ and a new one for Pomeron-Pomeron-Higgs vertex $V_H$.  Again in a self consistent  holographic approach to high energy scattering, one must work at  large but finite $\lambda$ where the Pomeron intercept is of the order $j_0\simeq 1.3$. As in the case of elastic scattering, the vertices will be evaluated at $\lambda = \infty$, unchanged from that calculated in the supergravity limit.  An explicit form for this amplitude will be given in Sec.~\ref{sec:BtoB}.  Here we focus on understanding the new vertex for Pomeron-Pomeron-Higgs fusion.
However a critical issue not address is the proton impact factors coupling to the Pomeron kernel.  Instead we assume a crude  phenomenological modification of AdS wave function for a  typical  glueball state as discussed in Sec.~\ref{sec:strategy}. For example we have found a surprisingly good  fits to HERA data for DIS at small  x 
by approximating the proton as fixed wave-function at the IR boundary.  We anticipate the  need to study this more seriously in the context of global fits to many diffractive processes that are more sensitive probes of this AdS  proton-proton-Pomeron vertex but it may well be that to a first approximation that the proton as seen by the Pomeron at large $N_c$ does appear to be very much like a spherical  glueball.

The key to our diffractive Higgs analysis is the recognition that, after integrating out the heavy quark loop, an external Higgs field   couples effectively  to the gluon Lagrangian density, $Tr[F^2]$,    which  by the AdS/CFT correspondence  is the source of the dilaton field at the boundary of AdS space.    Two particularly useful papers for our diffractive Higgs analysis are one by Herzog, Paik, Strassler and Thompson~\cite{Herzog:2008mu} on holographic double diffractive production of the scalar glueball and a second by Hong, Yoon and Strassler~\cite{Hong:2005np,Hong:2004sa} on the AdS/CFT vector form factor.  We will show that the double-diffractive Higgs production vertex, $V_H$, essentially  involves Pomeron-Pomeron fusion  producing a dilaton in the bulk of AdS  which propagates to the boundary via time-like  AdS scalar form factor.  However for this mechanism to work  an important new feature in double diffractive Higgs production, not emphasized in \cite{Herzog:2008mu}, is the need for conformal breaking in the bulk of the AdS space.  Without this conformal breaking in the bulk, the leading order  Pomeron-Pomeron dilaton vertex would vanish. Of course QCD is not a scale invariant theory so any model of AdS/QCD must include  some deformation of the AdS geometry.    When scale invariance is broken, one expects a non-vanishing vev for the gluon Lagrangian density, $F^2$.  More generally, we will be interested in correlators involving a single $F^2$, together with  any number of  stress-energy tensor $T_{\mu\nu}$, e.g.,  $\langle F^2(x) T_{\mu\nu}(y)T_{\mu\nu}(y')\cdots\rangle$. These correlators   can  be evaluated at strong coupling through the use of Witten diagrams involving the graviton-graviton-dilaton coupling in the bulk.   In a strictly conformal theory, scale invariance holds and all these correlators would vanish, corresponding to having  a vanishing graviton-graviton-dilaton vertex in the bulk.  This would in turn lead to a vanishing Higgs production vertex, $V_H=0$. That is, under such a scenario, central double-diffractive Higgs production would be suppressed at high energy.

Symmetry breaking effect in AdS/CFT correspondence has been studied in the past mostly using a near-boundary analysis~\cite{Witten:1998qj,Gubser:1998bc,Balasubramanian:1998sn,Banks:1998dd,Klebanov:1999tb, Polchinski:2000uf,Skenderis:2002wp}.     For our present purpose, it is more profitable to address scale invariance breaking in terms of Witten diagrams in the bulk. The simplest model, which we will use
as a first approximation,  is to  terminate the space in the IR (small radial co-ordinate $r$) at
a hardwall for confinement. The normalizable   modes are now discrete, giving rise to 
a glueball spectrum~\cite{Brower:2008cy}.  However since the bulk metric is unaffected, scale breaking effects resides on the IR wall only; for a non-vanishing  graviton-graviton-dilaton coupling     in the bulk,  a second {\it ad hoc} conformal breaking mass parameter must be introduced. This is not completely surprising  due to the
lack of a self-consistent  string dual for $N_c = \infty$ QCD. Fortunately it is possible to  fix the overall Pomeron-Pomeron-dilaton coupling by
appealing to the AdS/CFT dictionary.  In the gauge description, it is  fixed in terms of glueball matrix element of the trace of the energy momentum tensor, following the argument of Kharzeev and Levin~\cite{Kharzeev:2000jwa} as explained in Sec.~\ref{sec:anomaly} and also  in Appendix~\ref{sec:Confinement}.  This matching condition between  strong coupling and weak coupling is
the best we can do in lieu of solution to long sought dual string to large $N_c$ QCD.

\begin{figure}[bthp]
\begin{center}
\includegraphics[width = 0.7\textwidth]{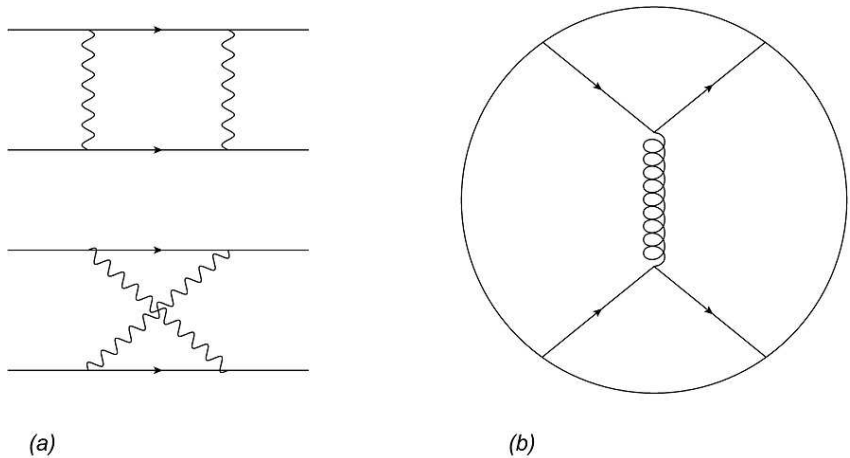}
\end{center}
\caption{Comparison between the extreme weak and strong coupling Pomerons:  (a) the 2 gluon exchange Low-Nussinov Pomeron at $g^2 N_c =0$  with intercept $j_0 = 1$ and  (b) the extreme strong coupling  Witten diagram of AdS-graviton ``Pomeron'' at $g^2 N_c  = \infty $ with intercept $j_0 = 2$.}
\label{fig:extreme}
\end{figure}

With the advent of the AdS/CFT correspondence,  we are now
able to understand diffractive Higgs production  starting from the extreme strong coupling limit.
In a sense our approach   is closer to production by the soft Pomeron, which is also an intrinsically non-perturbative
treatment,  used extensively by Donnachie and Landshoff and others to
parametrize high energy diffractive hadronic
process~\cite{DL,Donnachie:1998gm,Capella:1992yb}.
However holographic dual picture 
has the advantage of a unified soft and hard diffractive
mechanism. In the extreme strong coupling  limit the AdS/CFT dictionary
maps gauge theory into classical gravity in Anti-de Sitter space.
The leading correction to the strong coupling singularity~\cite{Brower:2006ea,Brower:2007qh,Brower:2007xg} for the the ${\cal N} = 4$ Super Yang Mills  is at 
\begin{equation}
j_0(\lambda) = 2 - 2 /\sqrt{g^2N_c}    \; ,
\label{eq:BPST-intercept}
\end{equation}
moving down from the bare graviton at  $J =2$, Fig. \ref{fig:extreme}(b), where $\lambda=g^2N_c$.    Comparing with weak coupling, we note
that in weak coupling it starts instead as the two gluon exchange at
$J = 2 - 1 = 0$ at $\lambda = 0$, Fig. \ref{fig:extreme}(a), and summing the BFKL ladder moves  to
\be
j_0(\lambda) = 1 + ( \ln 2/\pi^2) \; g^2N_c
\label{eq:BFKL-intercept}
\ee
to first order~\cite{Fadin:1975cb,Kuraev:1976ge,Kuraev:1977fs,Balitsky:1978ic}. To this order both weak and strong calculation are  consequences of
a leading order in  $1/N_c$ and conformal  approximation to Yang Mills theory. 
  Phenomenology for high energy cross sections  suggest $j_0 \simeq 1.3$ for the bare Pomeron intercept, 
squarely in the cross over region suggesting  both weak and strong coupling method
may be useful  to developing a reliable phenomenological ansatz for diffractive scattering 
in QCD. 

 We restrict ourselves in this paper primarily to the formulation of
the holographic amplitude with a  detailed analysis of the Pomeron-Pomeron Higgs
production vertex in a dual approach with scale invariance breaking.
We discuss   methods for developing a phenomenology which allows the 
inclusion of  the eikonal
corrections as well as the calibration by other diffractive processes for
elastic scattering, deep inelastic scattering and $t \bar t$ production.  This is not a
full phenomenological analysis but a first critical ingredient. In a
subsequent paper, we will use this formalism to preform a
self-consistent analysis of elastic diffractive scattering, deep
inelastic scattering and central diffractive heavy quark jets. Just as
in a more conventional weak coupling perturbative approach, these will be required to constrain
the parameters in the proton impact factors (or unintegrated gluon parton
distributions), and the Higgs diffractive vertex~\cite{Martin:2011gi}.  Although encouraging studies of baryons in holographic QCD have been carried out~\cite{Hashimoto:2010je,Hashimoto:2008zw,Yi:2008zz,Domokos:2010ma,Domokos:2009hm}, reliable calculations for meson- and nucleon-Pomeron vertex functions remain elusive.  Full
{\em ab initio} calculations are simply not possible at present even
in the $1/N_c$ expansion due to the lack of precise AdS dual to
QCD. Instead we must model properties of QCD by deforming the $AdS^5$
background metric to model the non-conformal consequences of
confinement and asymptotic freedom. Nonetheless consistency with a
full range of diffractive amplitudes is expected to lead to increasingly
useful predictions.  

This paper is organized as follows. In Sec  \ref{sec:Ingredient}, we give a
narrative for diffractive scattering and double diffractive Higgs production
from the AdS/CFT strong coupling view point. The goal is to
itemize the assumptions leading to our analysis. In Sec. \ref{sec:prelim},
we review the kinematics in the diffractive high energy limit
and give details on the building blocks required
to develop a model for holographic description of diffractive Higgs production. 
Sec.~\ref {sec:PPfusion}     deals
with the kinematic aspects of the new vertex $V_H$ for Pomeron-Pomeron fusion into
the dilaton, while  leaving to  Appendix~\ref{sec:Confinement}   a more detailed discussion on models with confinement deformation required for scale invariance breaking and for double-diffractive Higgs production.       Sec.~\ref{sec:strategy}  presents the normalization of double Pomeron Higgs production
amplitude by extrapolation to the tensor glueball on the Pomeron trajectory. This in principle completes the specification of the Higgs production amplitude, (\ref{eq:adsDoublePomeronHiggs}). In Sec.~\ref{sec:guess}, we provide  a phenomenological treatment under a simple-pole approximation,  leading to an estimate for  the central diffraction Higgs production cross section of $.8\sim 1.2  \;\; {\rm pbarn}$.  This is an over-estimate since it is arrived at without taking into account the absorptive correction, e.g., ``survival probability", which can lead to a central production cross section in the femtobarn range.
In Sec \ref{sec:discusion}, we conclude with comments on this and further corrections
needed to make a more reliable prediction.

\newpage
\section{Holographic Model for Diffractive Higgs Production}
\label{sec:Ingredient}

Diffractive scattering  and the notion of a Pomeron has always been an elusive object in QCD, often  defined
in a circular fashion as that which dominates high energy hadronic
scattering. In the large $N_c$ limit, there is a more precise
definition of the ``bare Pomeron''.  In leading order of the $1/N_c$
expansion at fixed 't Hooft coupling $\lambda = g^2N_c$, diffraction is
given peturbatively by the exchange of a network of gluons with the
topology of a cylinder, corresponding in a confining theory to the
t-channel exchange of closed strings for glueball states. Unitarity
imposes correction to the ``bare Pomeron'' in higher order in $1/N_c$:
(i) by adding closed quark loops to the cylinder, leading to
$q\overline q$ pairs or multi-hadron production via the optical
theorem dominated by low mass pions, kaon etc and (ii) by multiple
exchange of the Pomeron which includes the eikonal corrections (or
survival probability) and triple-Pomeron and higher order corrections
in a Reggeon calculus etc.  As discussed in the Introduction, the advent of the AdS/CFT correspondence has provided a firm framework for a non-perturbative treatment.

To arrive at a picture of the bare Pomeron it is useful to consider its form in both
weak and strong coupling. 
 Diffractive scattering in QCD has been explored extensively in the past from a perturbative approach, where, in the lowest order, it can be modeled as color singlet two-gluon exchange (or Low-Nussinov Pomeron) given in Fig.~\ref{fig:extreme}a  
and later  as a two Reggeized gluon ladder diagram (or the BFKL Pomeron) to first
order in the 't Hooft coupling $g^2 N_c$ and all orders $g^2 N_c
\log(s)$, with a BFKL intercept $j_0$ above unity given by (\ref{eq:BFKL-intercept}).   
An elastic amplitude $A(s,t)$ now grows with  a non-integer power as $s^{j_0}$, at $t$ fixed.

While the use of Regge poles to model  the Pomeron of non-perturbative QCD has
a long history, a more mathematically explicit picture arises with the conjecture  by Maldacena of an exact  equivalence between IIB super string theory on $AdS_5 \times S_5$ and
${\cal N} = 4$ SUSY Yang Mills theory that  holds true for all $N_c$ at fixed $\lambda$. The
$1/N_c$ expansion is string perturbation theory. We
presume (or hope) that this string/gauge duality also holds 
for pure Yang Mills theory (i.e. QCD) although no construction has
been found.    Thus the bare Pomeron is the
cylinder or closed string of the dual string.  In practice detailed calculation are based leading order at large  $N_c$ and  fixed 
't Hooft coupling $\lambda = g^2N_c$ which maps planar Yang Mills theory  to a free closed string theory, followed by a strong coupling expansion $\lambda \rightarrow \infty$ which
reduces  the strings to point like object for the classical solution of supergravity in 
an AdS like background.

It is instructive to plot the Pomeron intercept $j_0(\lambda)$ for both strong and weak
coupling as we have in Fig.~\ref{fig:N4j0} below.
\begin{figure}[bthp]
\begin{center}
\includegraphics[width = 3.7in]{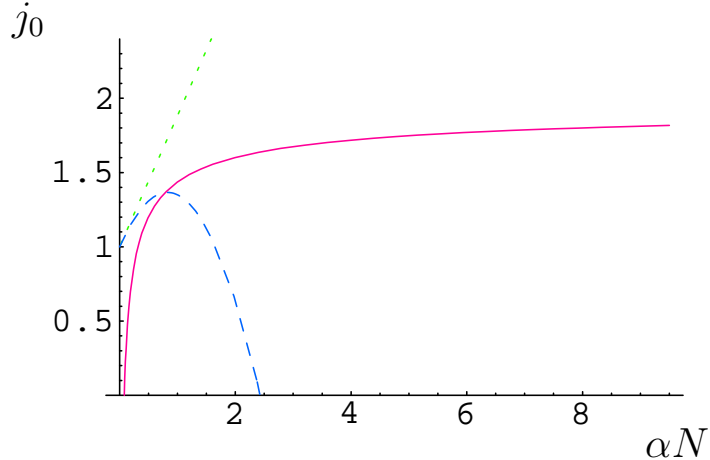}
\end{center}
\caption{In ${\cal N} = 4$ Yang-Mills theory,
the weak- and strong-coupling calculations of the
position $j_0$ of the leading singularity for $t\leq 0$,
as a function of $\alpha N = g^2 N_c/4 \pi$.
Shown are the leading-order BFKL calculation (dotted), the
next-to-leading-order calculation (dashed), and the strong-coupling
calculation of this paper (solid).  Note the latter two
can be reasonably interpolated. }
\label{fig:N4j0}
\end{figure}
 Perhaps it is an accident but the intersection of  the strong coupling curve in Fig.~\ref{fig:N4j0}
and the BFKL intercept  to second order occurs near to the phenomenological estimates, $j_0 \simeq 1.3$,  of the intercept  for QCD, suggesting that {\it the physics of diffractive scattering is roughly in the cross over region between
strong and weak coupling. }

We conclude this section with a few remarks on the framework we are
using to develop our strong coupling AdS/QCD model of diffraction. Readers
familiar with holographic QCD models may wish to skip this recapitulation. 
\begin{itemize}
\item{AdS Gravity and Confinement:}

In order to provide a particle interpretation, the basic framework for us is the holographic approximation to dual
QCD with confinement deformation~\cite{Polchinski:2001tt,Csaki:2008dt,Csaki:2006ji,Karch:2006pv,Erlich:2005qh,deTeramond:2011yi}.
At infinitely strong 't Hooft coupling, $g^2 N_c \rightarrow
\infty$,  the dual description, 
\begin{equation}
S=\frac{1}{2\kappa^2} \int d^5 x \sqrt g \Big[-{\cal R} - V(\phi) + \frac{1}{2} g^{MN} \dd_M\phi \dd_N \phi + \cdots \Big]\; ,
\label{eq:action}
\end{equation}
is assumed to be 5d  gravity coupled to dilaton field with a classical
vacuum approximated by $AdS_5 \times Y_5$ geometry in the bulk.  In principle we should solve the classical equation to define the background metric,
\begin{equation}
ds^2 = e^{2A(z)}[-dx^+ dx^- + dx_\perp dx_\perp + dz dz] + ds^2(Y^5)
\label{eq:nearAdS}
\end{equation}
Once the background geometry is known, expanding the action $S$ to quadratic order in metric fluctuation $h_{MN}$, the graviton kernel, $\widetilde {\cal K}_G$, can be found, (see  (\ref{eq:graviton}) for an explicit representation.).

 Of course there is no completely satisfactory example for such a background for
QCD even at large $N_c$.  Fortunately
for high energy Higgs production the dominant fluctuations are  the
graviton/Pomeron field and dilaton that couples to the Higgs at the boundary which are less sensitive to the details of confinement deformation. 
The ellipsis in (\ref{eq:action}) represents other fields or branes from unknown short distances physics 
that survive the strong coupling limit. For example, for the ${\cal N} = 4$ Super Yang Mills,
the pure $AdS_5 \times S_5$, background ( $\exp[2A(z)] = R^2/z^2$) , requires a 5-form Ramond-Ramond flux in order  to introduce the  cosmological constant 
\begin{equation}
V(\phi) = -\frac{12}{R^2}\; .
\end{equation}
  At lower energy additional 
fields are needed, e.g.,  the Kalb-Ramond $B_{\mu \nu}$ field is required for the $C = -1$ odderon as noted in Ref.~\cite{Brower:2008cy}.

With confinement deformation, the $AdS$ space is effectively cutoff in the interior. Because of the ``cavity effect", both dilaton and the transverse-traceless metric become massive, leading to an infinite set of massive scalar and tensor glueballs respectively.  In particular, each glueball state can be described by a normalizable wave function $\Phi(z)$ in $AdS$. The weight factor   $\Phi_{ij}$ in the respective factorized representation for the elastic and Higgs amplitudes, (\ref{eq:adsPomeronScheme}) and (\ref{eq:adsDoublePomeronScheme}), is given by $\Phi_{ij}(z) = e^{-2A(z)} \Phi_i(z)\Phi_j(z)$. In contrast, for amplitudes involving external currents, e.g., for DIS~\cite{Polchinski:2002jw,Brower:2010wf}, non-normalizable wave-functions will be used. 

\item{Correction to Strong Coupling in $1/\sqrt{\lambda}$:}

As pointed out earlier, taking into account $O(1/\sqrt \lambda)$ correction to the Graviton kernel, $ \widetilde{\cal K}_G$, one arrives at (\ref{eq:adsPomeronScheme}) and (\ref{eq:adsDoublePomeronScheme}) for elastic and diffractive Higgs production respectively. Here the Pomeron kernel, $ \widetilde{\cal K}_P$,  given explicitly in a $J$-plane representation,  (\ref{eq:graviton}), has hard components due to near conformality in the UV and soft
Regge behavior in the IR. 
We stress that this first order strong coupling correction corresponds to a stringy effect,
as  has been demonstrated in \cite{Brower:2006ea} by introducing a Pomeron world-sheet  vertex operator ${\cal V}_P$ while enforcing the on-shell condition,
\begin{equation}
(L_0 -1){\cal V}_P=(\bar L_0 -1){\cal V}_P =0 \; .
\end{equation}
As discussed in Ref.~\cite{Brower:2008cy}, this can also be carried out for the anti-symmetric field $B_{\mu\nu}$, leading to a description for Odderon in the strong coupling limit.

It is worth repeating that the importance of this stringy correction enters most importantly in the modification of the $s$-dependence of the Pomeron kernel, $\widetilde{\cal K}_P$, with $j_0$ moving from 2 to a phenomenological value close to 1.3. However, for wave-function in (\ref{eq:adsPomeronScheme}) and (\ref{eq:adsDoublePomeronScheme}), stringy corrections can be ignored.  With this understanding, the 2-to-2 glueball scattering amplitude,  (\ref{eq:adsPomeronScheme}),  written  in terms of AdS radial coordinate, becomes
\begin{equation}
A(s,t) = \int dz dz'\; \sqrt{-g(z)} \sqrt{-g(z')}\; \Phi_{13}(t,z)  \widetilde {\cal K}_P(s,t,z,z')    \Phi_{24}(t,z') \; ,
\label{eq:adsPomeron}
\end{equation}
A more explicit form  for the Pomeron kernel will be given in Appendix~\ref{sec:Confinement}. 

\item{Weak Coupling Higgs Production:}

In a
perturbative approach, often dubbed as ``hard Pomeron", Higgs production can be viewed as gluon
fusion in the central rapidity region~\cite{Martin:2009zze}.  A Higgs can be produced at central rapidity by the double Regge Higgs vertex through
a heavy quark loop which in lowest order is a simple gluon fusion process
as illustrated in Fig.~\ref{fig:ExtremeHiggs}a dominant for large parton x for the colliding gluons.
A more elaborate picture emerges as one tries to go  to the region of the softer (wee gluons) building up
double Regge regime. In addition to  the Pomeron exchange contribution in these models must
subsequently be reduced by large Sudakov correction at the Higgs
vertex and by so-called survival probability estimates for soft gluon
emission, not inconsistent with  the view of some that double diffractive Higgs
production should be intrinsically non-perturbative.

\begin{figure}[bthp]
\begin{center}
\includegraphics[width = 0.4\textwidth]{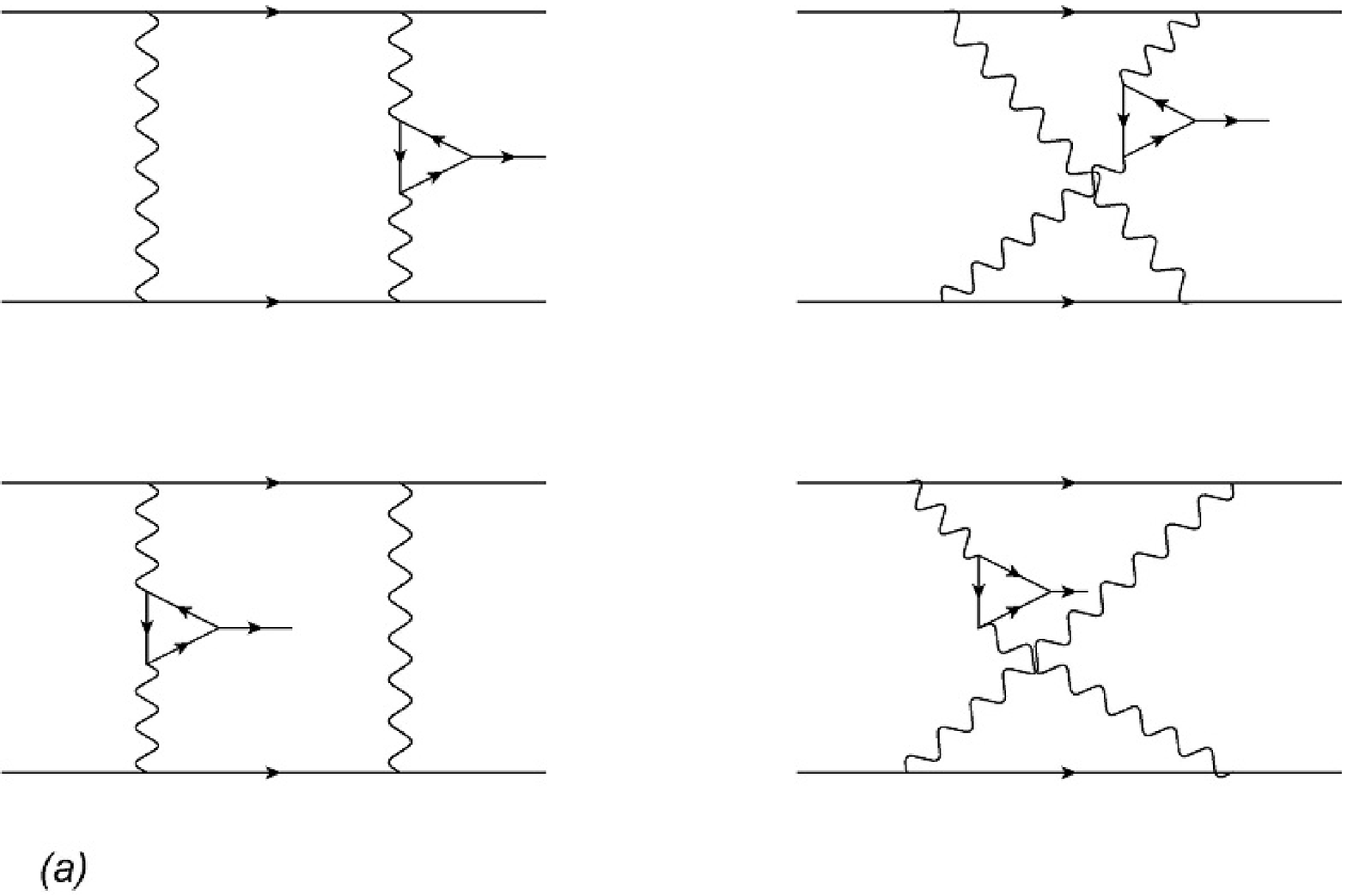}
\hskip 3cm
\includegraphics[width = 0.25\textwidth]{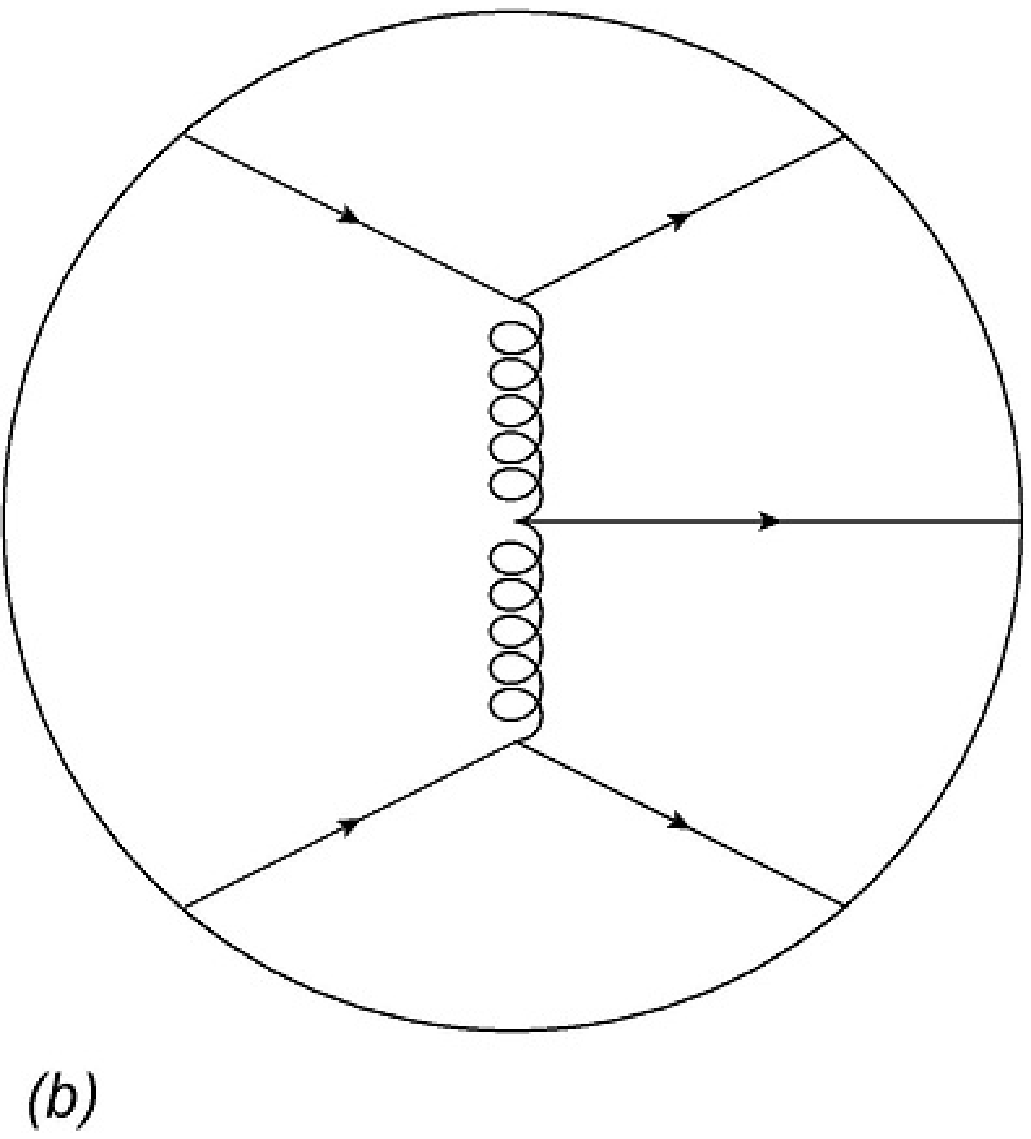}
\end{center}
\caption{(a) Higgs production by gluon fusion in the 2 gluon exchange Low-Nussinov Pomeron at $g^2 N_c
  \rightarrow 0$ vs (b) The  Witten diagram for Higgs production
by AdS graviton fusion at $1/g^2 N_c
  \rightarrow 0$. The graviton fusion is a source of
a bulk to boundary scalar the propagator for the heavy quark loop
at the boundary.}
\label{fig:ExtremeHiggs}
\end{figure}
\begin{figure}[bthp]
\begin{center}
\includegraphics[angle = 90,width = 0.5\textwidth]{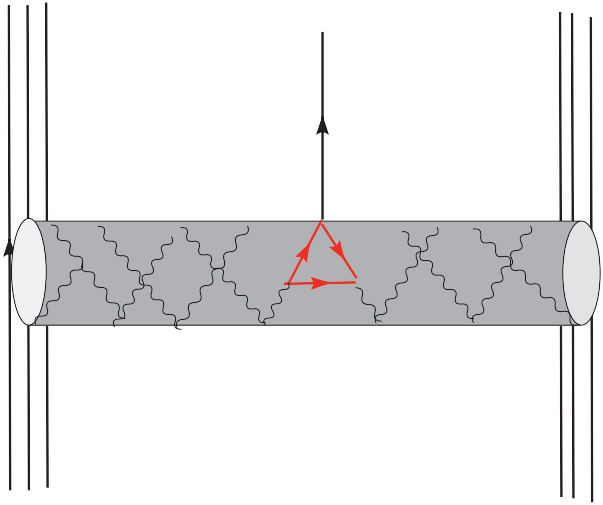}
\end{center}
\caption{Cylinder Diagram for large $N_c$ Higgs Production.}
\label{fig:cylindarHiggs}
\end{figure}
\newpage 
\item{Strong Coupling Higgs Production:}

In the large $N_c$ there are no quark loop in the bulk of AdS space and since
the Higgs in the Standard Model only couples to quarks  via the
Yukawa interactions there appears to be a problem with strong coupling Higgs production
in leading $1/N_c$.   Fortunately the solution to this is to follow the standard procedure in Higgs phenomenology, which is to
integrate out the quark field replacing the Higgs coupling to the gauge operator $Tr[F^2]$. 

Consider the Higgs coupling to quarks via a Yukawa coupling,  and, for simplicity we will assume it is dominated by the top quark. We will be more explicit in the next Section, and simply note here that, 
 after taking advantage of the scale separations between the QCD scale, i.e., the Higgs mass and the top quark mass,   $\Lambda_{QCD} \ll m_H \ll 2 m_t$, 
heavy quark decoupling allows one to replace the Yukawa coupling  by an effective interaction,
\begin{equation}
{\cal L} = \frac{\alpha_s g}{24 \pi M_W} F^a_{\mu\nu}F^{a \mu \nu} \phi_H
\end{equation}
by evaluating the two gluon Higgs triangle graph in leading order $O(M_H/m_t)$. 
Now the AdS/CFT dictionary simply requires that this be the source in the UV of the AdS dilaton field.  It follows, effectively, for Higgs production, we are required to work with a five-point amplitude, one of the external leg involves a scalar dilaton current coupling to $Tr[F^2]$.  For diffractive Higgs production, in the supergravity limit, the Higgs vertex $V_H$ is given by a two-graviton-dilaton coupling, Fig.~\ref{fig:ExtremeHiggs}b. After taking into account finite $\lambda$ correction, the leading order diagram at large $N_c$ can be schematically represented in Fig.~\ref{fig:cylindarHiggs}, with each of the left- and right-cylinders representing a BPST Pomeron.

\item{Conformal Symmetry Breaking:}

  Of course QCD, even at $N_c = \infty$, is not a conformal theory. Conformal symmetry breaking (or ``dimensional transmutation'' in the colorful language of Sidney Coleman) is ultimately tied to confinement in the IR and asymptotic freedom in the UV. A true QCD dual (or QCD string theory) would require an infinite number of (higher spin) fields in the bulk representation to correspond to fluctuations in the yet undiscovered world-sheet string theory for QCD.  All mass scales (for quarkless large $N_c$ QCD) are related and the coupling $\lambda = g^2 N_c$ is not a free parameter.  Fortunately  at high energy, these details are non-essential. For our purposes an adequate phenomenological AdS  dual
to QCD  requires only two features: (1) an IR deformation, which for simplicity we take as hard-wall cut-off beyond $z = 1/\Lambda_{QCD}$, to give confinement and  a linear static quark potential at large distances and (2) a slow deformation in the UV ($z \rightarrow 0$) to model the logarithmic running for asymptotic freedom.  Moreover scale breaking plays an even more essential role in the application to  diffractive Higgs production. 
In leading order in strong coupling, Pomeron-Pomeron fusion proceeds  through
a graviton-graviton-dilaton vertex,  which is lacking in AdS gravity action (\ref{eq:action}) if scale invariance is maintained. This is demonstrated more explicitly in Appendix~\ref{sec:Confinement}. This
vertex, $M^2 \phi h_{\mu\nu} h^{\mu \nu}$, itself requires a scale breaking mass parameter.   However we will show, based on the AdS/CFT dictionary, that this new mass parameter can be fixed by computing a  matrix element of the trace of the energy momentum tensor for the tensor glueball on the Pomeron trajectory  following the argument of Kharzeev and Levin~\cite{Kharzeev:2000jwa}. The overall rate of Higgs production is not a free parameter of our AdS/QCD model. Still all strong coupling QCD  duals to date fail to relate the IR confinement scale to the UV scale
 (or to  $\lambda$) so  a least one extra  mass needs to  be fixed phenomenologically.
\end{itemize}

\newpage

\section{Kinematics and Regge Analysis of Building Blocks}

\label{sec:prelim}

The natural coordinates for high energy scattering in warped $AdS^5$ space are given in the Poincare patch in lightcone coordinates,
\be
ds^2 = e^{2A(z)}[-dx^+ dx^- + dx_\perp dx_\perp + dz dz] \; ,
\ee
where the incoming particles are directed near to the light cones $x_\pm = x^0 \pm x^z \simeq 0$ and the transverse impact parameters
are extended to 3-dimensions, ${ \bf b}=(x_\perp,z) = (x_1, x_2,z)$, the traditional 2-d transverse space, ${\bf x}_\perp = (x_1,x_2)$, plus the ``radial'' coordinate $z = R^2/r$.
Here we relate  this picture to the standard Mandelstam coordinates for 2-to-2 and 2-to-3 amplitudes in the Regge limit.   We also 
introduce the analytic $J$-plane where the AdS  Pomeron kernels  take their simplest form.

\subsection{Elastic Diffractive Scattering}

For high  energy elastic (nucleon) scattering,
\begin{equation}
 p(k_1) + p(k_2) \rightarrow p(k_3) + p(k_4) \; ,
\end{equation}
the exact connection between light-cone coordinates and
Mandelstam invariants are simplest in the brick-wall frame where $\bf k_{1 \perp} = -\bf k_{3 \perp} ={\bf q_\perp}/2$.  In terms of the total rapidity $y$, the invariants
are   $s = (k_1 + k_2)^2 \equiv 4 m_\perp^2 \cosh y$, $m^2_\perp=m^2+ {\bf q_\perp}^2$ and  $t = (k_1 - k_3)^2 =   - {\bf q_\perp}^2$. The
two component transverse  momentum vector $ {\bf q_\perp} = (q_1, q_2)$ is the Fourier
transform of the impact co-ordinates $x^1, x^2$. 
In the super-gravity limit where $\lambda \rightarrow \infty$, the one-graviton diagram grows as $s^2$, (see (\ref{eq:graviton2})), and the one-graviton kernel is given by 
\begin{equation}
\widetilde {\cal K}_G(s,t,z,z') = s^2 ( zz'/R^2)^2 \widetilde G_2(z,z',t)
\label{eq:graviton}
\end{equation}
where
\be
\widetilde G_2(z,z',t = - {\bf q}^2_\perp)  = 
\int \frac{d^2x_\perp}{ 4 \pi^2}  
e^{\textstyle i {\bf q}_\perp \cdot ( {\bf x}_\perp - {\bf x'}_\perp)}  G_2(z,z', {\bf x}_\perp - {\bf x'}_\perp)
\ee
In the conformal limit,    $\widetilde G_2$ is simply the ``massless" $AdS_5$ propagator in a momentum representation, satisfying a simply differential equation
\begin{equation}
\left( -z\partial_z z\partial_z + 4 - z^2 t\right) \widetilde  G_2(z,z';t) = z \; \delta(z-z')  
\end{equation}
At   finite $\lambda$, it has been shown in Ref. \cite{Brower:2006ea} that, due to curvature of AdS, the effective spin of a graviton exchange is lowered from 2 to $j_0=2-2/\sqrt \lambda$. As such it is necessary to adopt a $J$-plane formalism where the Pomeron kernel $\widetilde {\cal K}_P$  is given by an inverse Mellin transform~\footnote{For ease  of writing, we have introduced $\widetilde {\cal K}_P$ in this paper, which is simply related to ${\cal K}_P$ used in Refs. \cite{Brower:2006ea, Brower:2007qh,Brower:2007xg,Brower:2010wf} by an AdS factor $\widetilde {\cal K}_P=  e^{(2A(z)+2A(z'))} {\cal K}_P$. This notation is more convenient in order to accommodate the use of impact factor $\Phi_{ij}= e^{-2A}\Phi_i\Phi_j$. For simplicity, we have also used through out, for this purpose, $e^{-2A}= (z/R)^2.$}, 
\begin{equation}
\widetilde{\cal K}_P(s,t,z,z') =
-  \int_{-i\infty}^{i\infty} \frac{dj}{2\pi i} (
\alpha'  \widehat s)^{j} \frac{1 + e^{-i \pi j}}{\sin\pi j}   \widetilde G_j(t,z,z') \; .
\label{eq:PomeronKernel}
\end{equation}
with $\widehat s= zz's/R^2$.  When conformal invariance is maintained,   $\widetilde G_j(t,z,z')$ satisfies a $J$-dependent ``massive" $AdS_5$ differential equation
\bea
\left( -z\partial_z z\partial_z + (2\sqrt{\lambda})(j-j_0) - z^2 t\right) \widetilde  G_j(z,z';t) &=& z \; \delta(z-z')  
\label{eq:tensor}
\eea
We remind the reader that the Regge $J$-plane is  the conjugate variable to the light-cone boost operator: $\hat H = M_{+-}$ which in  principle
provides an exact one to  one map for amplitudes using the Laplace/Mellin transform. To accomplish this one must transform separately   contributions from exchanges of definite charge conjugation, $C=\pm 1$, or, more precisely, Regge contributions with a definite signature.  The leading singularities for $C = \pm 1$ are referred to as the
Pomeron and Odderon~\cite{Brower:2008cy} respectively. For the
closed string theory these exchanges are
associated with the graviton and Kalb-Ramond fields
respectively~\cite{Brower:2008cy}.  Applying this analysis to the AdS Pomeron amplitude, as detailed in Ref. \cite{Brower:2006ea}, the elastic amplitude at high energy  can again be represented schematically in a factorized form, Eq. (\ref{eq:adsPomeronScheme}).

The salient new element of the Regge formulation in the AdS/CFT description is the extra radial (or 5th) coordinate ($z = R^2/r$). The variable 
conjugate  to $\log (z/R)$ is $\nu$, which, up to a constant shift, is the conformal dimension, which in turns allows an inverse-Mellin representation~\cite{Brower:2006ea}. It follows from Eq. (\ref{eq:tensor}) above that conformal Pomeron at $t = 0$, in a double-Mellin representation,  is a simple pole in the $J-\nu$ plane, 
\be
 \widetilde G_j(\nu, t = 0)  \sim    \frac{1}{(2\sqrt{\lambda})(j-j_0) +\nu^2  }
\ee
For non-zero $t$ the full expression is given Eq.~(\ref{eq:adsPompropagatorJ}) of 
Appendix \ref{sec:Confinement}.  The coordinate z plays the role of ``virtuality'' in the partonic language of Yang Mills theory.   As explained in \cite{Brower:2007qh,Brower:2007xg,Cornalba:2006xm,Cornalba:2006xk,Cornalba:2007fs,Cornalba:2007zb}, in the conformal limit, in an impact representation  at high energy, it leads to a transverse $AdS^3$ (or 3-d Hyperbolic space $H_3$), and a corresponding simpler expression for the Pomeron kernel, (\ref{eq:conformalkernel}).

\subsection{Double Regge Analysis}

The Regge analysis for 2-to-3 amplitude follows the same path as for the elastic amplitude but is considerably more subtle. The double diffractive Higgs production amplitude, $A(s,s_1,s_2,t_1,t_2)$,
\begin{equation}
 p(k_1) + p(k_2) \rightarrow p(k_3)+  H(q) + p(k_4)
\end{equation}
has 5 Mandelstam invariants.   Again  we can express 
the invariants, $s = (k_1 + k_2)^2$, $s_1 = (k_3 +q)^2, t_1 = (k_1 - k_3)^2, s_2 = (k_4- q)^2, t_2 = (k_2 -k_4)^2$  in terms of light cone coordinates by choosing an 
appropriate frame:
\begin{align}
k_1 &= (m_1 e^{y/2},m_1 e^{- y/2},0_\perp)   \qquad ,   & k_2
&= (m_2 e^{- y/2}, m_2 e^{y/2},0_\perp)  \nonumber \\
k_3 &=  (m_{3\perp} e^{y_3},m_{3 \perp} e^{-y_3},-q_{1 \perp })
\qquad , &k_4 &= (m_{4
  \perp} e^{- y_4},m_{4 \perp} e^{y_4},- q_{2\perp})
\end{align}
where the transverse mass  is $m_\perp = m^2 + k^2_\perp $ for any on shell state $k^2_i = m^2$.  The momentum for the central particle, $q= k_1+k_2-k_3-k_4$   be parametrized by  $q = (m_{H\perp} e^{y_H},m_{H\perp}e^{-y_H},q_{H\perp})$ with transverse
mass  $m_{H \perp}$,
\begin{equation}
m^2_{H \perp} =  m^2_H + q^2_\perp \; . 
\end{equation}

The double Regge limit is $|y_3-y_H|  \sim \log(s_1) \rightarrow \infty$,  $|y_4+y_H|  \sim \log(s_2) \rightarrow \infty$ at
fixed momentum transfers $t_i \simeq - k^2_{i\perp}$ and   ${\bf q}_\perp^2$.  In this limit, one also has $y_3\simeq -y_4\simeq y/2$. 
The Higgs momentum components are fixed by energy-momentum conservation, e.g.,\begin{equation}
y_H = \frac{1}{2}[\log(s_1/ s_2) - \log(m_{3\perp}/m_{4\perp})] \; .
\end{equation}
The transverse
mass is specified  by the  relation,
\begin{equation}
\kappa = \frac{s_1 s_2}{s} \simeq m^2_H + q^2_\perp = m^2_{H \perp}\; , \label{eq:kappa}
\end{equation}
referred to as the ``kappa'' variable for the Regge-Regge particle vertex,
$V(t_t,t_2,\kappa)$. It is natural to introduce the two outgoing rapidity gaps
separating the Higgs as
\begin{equation}
\Delta y_1 = y_3 - y_H  \quad , \quad \Delta y_2 = y_4 + y_H \; .
\end{equation}
The light-cone parametrization provides exact change of coordinates: $s,s_i,t_i \rightarrow  \Delta y_i, q^2_{i\perp}$. The double diffractive production limit is characterized by small transverse momenta, $t_i \simeq - q^2_{i\perp }$ and large rapidity gaps, $\Delta y_1 \simeq \log(s_1/(m_{3\perp } m_{H\perp}))$ and $\Delta y_2 \simeq \log(s_2/(m_{4\perp } m_{H\perp}))$.
\begin{figure}[bthp]
\begin{center}
\includegraphics[width = 0.4\textwidth]{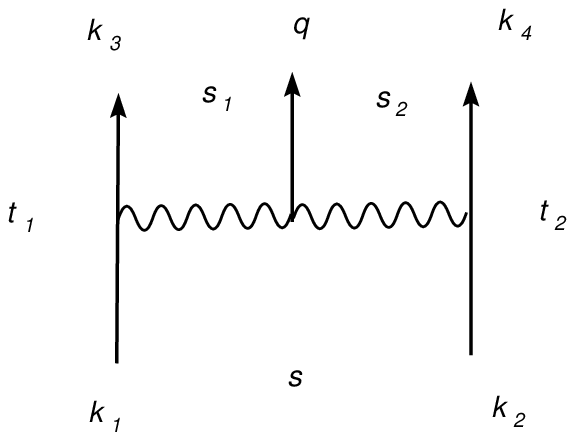}
\hskip 2 cm
\includegraphics[width = 0.4\textwidth]{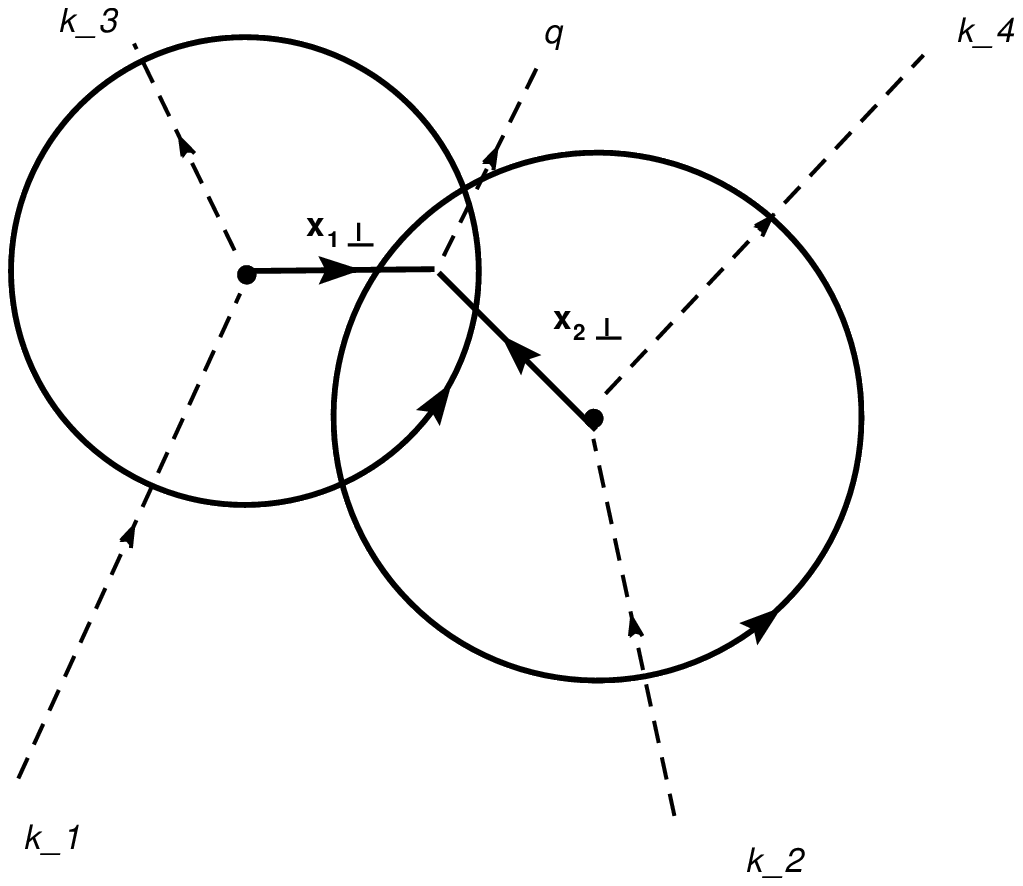}
\end{center}
\caption{On the left the double Regge kinematics for $p(k_1) + p(k_2)
  \rightarrow p(k_3) + H(q) + p(k_4)$. At fixed center of mass energy
  $\sqrt{s}$, the final state configuration determined by 4 invariants
  $t_1,s_1, t_2,s_2$. On the right, the impact parameter picture with
  the Higgs at $x_{H\perp}$ at a separations  $\Delta x_{i\perp} =
  x_{H\perp} - x_{i\perp}$ of the two incoming particles. External momenta are for the $\pm$ light-cone components only. }
\label{fig:regge5}
\end{figure}

In the double Regge limit the only ``$\kappa$'' dependence occurs in the
Regge-Regge Higgs vertex, $V(t_t,t_2,\kappa)$. Note also since $\kappa =
m^2_H + q^2_{1\perp} + q^2_{2\perp} + 2 q_{1\perp} \cdot q_{2\perp}$, $\kappa$ is solely responsible for the  so-called Toller angular dependence in $q_{1\perp} \cdot q_{2\perp}$
between the two protons scattering planes (see Fig.~\ref{fig:regge5}).
 The key to a double-Regge expansion is factorization in the $J$-plane. It is possible to obtain the full amplitude via a double-inverse-Mellin transform.   We refer to the literature
for further discussion of this formalism for   some of the subtleties involved~\cite{Brower:1974yv}.  Most of these  can be neglected 
due to the high mass of the Higgs at the central vertex. Consequently,
with this {\it proviso}, we arrive at the amplitude for Higgs production expressed in terms of two Pomeron kernels, $ \widetilde{\cal K}_P(s_1,t_1,z_1,z_1')$ and  $\widetilde{\cal K}_P(s_2,t_2,z_2',z_2)$, together with a Higgs vertex, $V_H( t_1,t_2, \kappa, z_1',z_2') $.  Finally we
are able to give an explicit form to the symbolic expression for
 the double-Pomeron amplitude in Eq. \ref{eq:adsDoublePomeronScheme} of
the Introduction:
 \bea
A(s,s_1, s_2, t_1, t_2) &\simeq &  \int dz_1 dz_1' dz_2'  dz_2\; \sqrt{-g(z_1)}\sqrt{{-g(z'_1)}}\sqrt{-g(z'_2)}\sqrt{-g(z_2)}\;   \\
&\times&   \Phi_{13} (t_1,z_1) \widetilde{\cal K}_P(s_1,t_1,z_1,z_1')  V_H( t_1,t_2, \kappa, z_1',z_2') \widetilde {\cal K}_P(s_2,t_2,z_2',z_2) \Phi_{24}(t_2,z_2)  \;. \nonumber 
\label{eq:adsDoublePomeron}
\eea
In the above expression,  impact factors,
$\Phi_{13}(t_1,z_1)$ and $\Phi_{24}(t_2,z_2)$ are identical to those
entering the elastic amplitude (\ref{eq:adsPomeron}).
This represents a generalization of Eq. (\ref{eq:adsPomeron})
for elastic scattering to central diffractive Higgs production. The
only new component is the
Pomeron-Pomeron-Higgs vertex, $V_H( t_1,t_2, \kappa, z_1',z_2')$, which in general allows non-local interaction in $AdS_3$, e.g., its dependence on $z_1'$ and $z_2'$.

Just as the case of 2-to-2 scattering, it is often simpler to think of scattering in terms of transverse coordinates ${ \bf b}=(x_\perp,z)$. Instead of the vertex $V_H( t_1,t_2, \kappa, z_1',z_2')$ in momentum space, one can work directly with $V_H (x'_{1\perp} -x_{H\perp},z'_1;x'_{2\perp}
-x_{H\perp},z'_2)$. The $\kappa$-dependence enters through the possible dependence on the  angular correlation between the two proton's scattering planes,  $ x'_{1\perp} -x_{H\perp}$ and $x'_{2\perp}
-x_{H\perp}$ or the scalar product: $  (x'_{1\perp} -x_{H\perp}) \cdot (x'_{2\perp}
-x_{H\perp})$.  As we will see, for Higgs production, due to its large mass, this dependence is suppressed and we will choose to drop it. Indeed, (\ref{eq:adsDoublePomeron}) would take on a more complicated form if non-trivial $O(1/m_H^2)$ dependence on $\kappa$ turns out to be important. We will not address this issue here.
\newpage

\section{Pomeron-Pomeron fusion Vertex}
\label{sec:PPfusion}

We now turn to a detailed discussion of the double diffractive Higgs vertex, $V_H$.  So
far our discussion  of double-Pomeron exchange (\ref{eq:adsDoublePomeronScheme}) applies equally well  both to  diffractive glueball production and  to Higgs production.  The
situation is similar  to  how we converted the amplitude for  proton-proton (p-p) elastic scattering to  electron-proton deep-inelastic scattering (e-p DIS).  For  DIS, we simply replaced  the  normalizable proton wave-functions in (\ref{eq:adsPomeron}) with non-normalizable counterparts appropriate for conserved external vector currents.    For
 glueball production, the vertex is proportional to a normalizable $AdS$ glueball wave-function, whereas for  Higgs production, the  vertex, $V_H$, requires a non-normalizable
bulk-to-boundary propagator, $K(q^2,z)$, appropriate for a scalar external
current. This section will explain how this transformation is done in detail.

\subsection{Higgs Vertex}

A Higgs scalar in the standard model~\cite{Englert:1964et,Higgs:1964pj,Guralnik:1964eu,Guralnik:2011zz,Higgs:1966ev}  couples exclusively to the quarks
via Yukawa coupling, which for simplicity we will assume is dominated
by the top quark,
 \begin{equation}
{\cal L} = - \frac{g}{2 M_W} m_t \; \bar t(x) t(x) \phi_H(x)\; .
\label{eq:HiggsCoupling2ttbar}
\end{equation}
If we assume a mass for Higgs that obeys 
\begin{equation}
 \Lambda_{QCD} \ll  m_H \ll 2 m_t \; ,
\end{equation}
we can use the conventional heavy quark approximation~\cite{Gunion:1989we} by integrating out the top quark loop replacing the Yukawa coupling by the effective coupling of Higgs field to the gluon operator,
\begin{equation}
{\cal L} =L(q^2) F^a_{\mu\nu}F^{a \mu \nu} \phi_H \; ,
\label{eq:effectiveHiggsCoupling}
\end{equation}
where
\begin{equation}
L_H\equiv L(-m_H^2)  \simeq  \frac{\alpha_s g}{24 \pi M_W} \; , 
\label{eq:LH}
\end{equation}
to leading order $O(m_H/m_t)$. In the dual AdS description at leading order in $1/N_c$ heavy quarks are external sources near to the boundary of AdS so this procedure just replaces the source $\bar t(x) t(x)$   by $F^2(x)$ at the boundary.  Consequently 
the vertex $V_H (x'_{1\perp} -x_{H\perp},z'_1;x'_{2\perp}
-x_{H\perp},z'_2)$ is now given  by the product of a   Pomeron-Pomeron-dilaton
vertex in the bulk and 
\begin{figure}[bthp]
\begin{center}
\includegraphics[width = 0.5\textwidth]{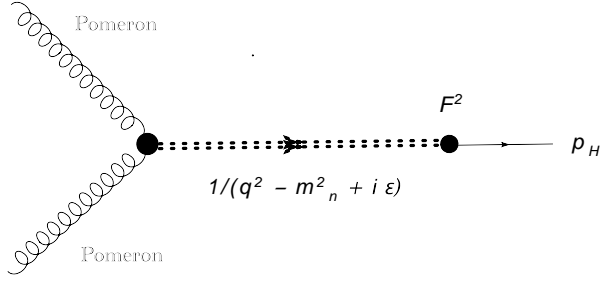}
\end{center}
\caption{Factorization of double diffractive vertex in strong coupling:
  From right to left:  $F^2(x)$-Higgs
vertex  the boundary of AdS ($z_H =0$),  the boundary to scalar propagator
saturated by glueballs
and, finally, the double Pomeron to scalar
vertex in the interior of AdS ($z_0 $).}
\label{fig:higgs_vertex}
\end{figure}
a scalar bulk-to-boundary propagator,
$K(x'_H - x_H,z)$ from the interior of $AdS_3$ at ${\bf b}_H=(x'_H,z)$ 
to the source $F^a_{\mu \nu}F^a_{\mu \nu}(x)$ at  $z_0 \rightarrow 0$.
In the most general form, the Pomeron-Pomeron-dilaton   vertex can be
 non-local where the two Pomerons ``fuse''  at  $z'_1$,  $z'_2$  into a dilaton at   $z$.  So our vertex would take the form, 
\begin{equation}
V_H(t_1,t_2, \kappa, z'_1,z'_2)=L_H \int d z {\cal V}(t_1,t_2,\kappa, z'_1,z'_2, z )K(q^2,z) \; ,
\end{equation}
evaluated on shell at $q^2=-M_H^2$.   However in the spirit of strong coupling,  we impose locality on the Pomeron-Pomeron scalar vertex in the bulk,
\begin{equation}
{\cal V}(t_1,t_2,\kappa, z'_1,z'_2, z ) =( {1/\sqrt{-g}})\delta(z'_1 - z) \delta(z'_2 -z) {\cal V}_L(t_1,t_2,\kappa, z )
\end{equation}
Therefore, the overall central Higgs vertex $V_H$  can be written as a product of several factors, 
\begin{equation}
V_H(t_1,t_2, \kappa, z,z'_2)=( {1/\sqrt{-g}}) \delta(z - z'_2)
L_H  {\cal V}_L(t_1,t_2,\kappa, z) K(q^2,z)
\label{eq:factorization}
\end{equation}
Of course this first step, taking the heavy quark limit, $ m_H \ll m_t$,  is simply the standard approximation
often used in Higgs production for both soft and hard gluon
fusion.
We can achieve further simplifications  by   taking advantage of the fact that $\Lambda_{QCD} \ll  m_H $. Before doing so, it is instructive to first re-consider the case of production of scalar glueballs.

\subsection{Pomeron-Pomeron-Glueball  Vertex}

In a treatment  with confinement deformation, the bulk-to-boundary propagator can be saturated by a set of scalar glueballs with increasing masses. The factorized structure, (\ref{eq:factorization}), can then be schematically 
represented by Fig. ~\ref{fig:higgs_vertex}.  It is therefore useful to first consider  double diffraction
vertex for scalar production of a massive glueball of mass $m_n$, $V_n(t_1,t_2,\kappa,z_1',z_2')$.  Here $\kappa$ is a standard double-Regge invariant, $\kappa = s_1s_2/s$, which we will return to shortly.
 In momentum space, this vertex can be expressed as
\bea
V_n(t_1,t_2, \kappa, z'_1,z'_2 )
&=& \int d z \; {\cal V}_n(t_1,t_2,\kappa, z'_1,z'_2, z) \;  \phi_n(z)
\eea
where we have  labeled scalar glueballs by $n$, with a wave-function $\phi_n(z)$ in $AdS$ and a coupling $f_n$ to two Pomerons. As noted above, the coupling ${\cal V}$ is  in general non-local in $AdS$, depending on $z'_1,z'_2, z$. However, in the super-gravity limit, scattering becomes local, as schematically represented by the Witten diagram, Fig.~\ref{fig:ExtremeHiggs}b, i.e., $z'_1=z'_2=z$.  In fact, in this limit, ${\cal V}_n(t_1,t_2,\kappa, z'_1,z'_2, z')\rightarrow {\rm constant}$, leading to %
\bea
V_n(t_1,t_2, \kappa, z'_1,z_2' )
&=&\delta(z_1'-z_2')    \; {\cal V}_n  \int d z\; \delta(z_1'-z) \; \phi_n(z)
\eea
 However for $\lambda$ finite, we should entertain non-trivial dependence on its arguments,
at least consistent with that which is seen in flat-space string theory. But we
can still assume locality in the AdS radial co-ordinate as advocated in Ref.~\cite {Hong:2005np}, leading to 
\bea
V_n(t_1,t_2, \kappa, z'_1,z_2' )
&\simeq & \delta(z_1'-z_2')  \int d z \delta(z_1'-z) {\cal V}_n(t_1,t_2,\kappa,  z) \; \phi_n(z).
\eea
 In a double-Regge limit, $\kappa$ is kinematically given by the transverse momentum by $\kappa=m_n^2+ q_\perp^2$.  In a frame where incoming particles 1 and 2 are
longitudinal, translation invariance leads to $ q_{\perp} = -( q_{3\perp} +
q_{4\perp} )$ and rotational invariance then allows the vertex to depend
only on magnitudes of $q_{3\perp} $ and $q_{4\perp} $, and on the relative
angle between them, e.g., $|q_{3\perp}|$, $|q_{4\perp}|$, and their vector dot-product $
q_{3\perp}\cdot q_{4\perp}$. Equivalently, since $\kappa=m_n^2+ q_\perp^2$, the
vertex can also be considered as a functions of $t_1= - q_{3\perp}^2$, $t_2=-q_{4\perp}^2$, and  $q_{\perp}^2$. Therefore, ${\cal V}(t_1,t_2,\kappa, z')$  provides  the most general information appropriate for  local Pomeron fusion in AdS.

In flat space the  dependence  of  ${\cal V}(t_1,t_2,\kappa, z)$ on $t_1,t_2,\kappa$ has
been studied extensively.   For an arbitrary external particle with spin, the vertex will also have
additional spin dependence. However, for our present purpose, we only
need to consider scalar production, with its mass $m_n^2$ allowed to
take on large values. Let us focus on the dependence of ${\cal V}_{flat}(t_1, t_2,\kappa)$ on the variable
$\kappa$.  The general analytic structure  in $\kappa$ is robust. For open-string, it is real for $\kappa<0$, with branch cut along the positive real axis, $\kappa\ge 0$.   At $\kappa=0$, one has
\begin{equation}
{\cal V}_{flat}(t_1, t_2,\kappa)\simeq (-\kappa^{-1})^{\alpha_1} G_1 +(-\kappa^{-1})^{\alpha_2} G_2
\label{eq:smallkappa}
\end{equation}
where $G_1$ and $G_2$ are regular at $\kappa=0$. This analytic structure~\footnote{For  related  recent studies, see \cite{Brower:2008ia} and references therein.},  (or
Steimann relations), is required to avoid overlapping discontinuities not present in 
planar diagrams on the one hand and in the Regge limit of open strings on the other.
 For closed strings, a similar analysis can also be carried out, with branch points at $\kappa=0$ and at $\kappa=\infty$,  extending to the entire real axis. Returning to the case of $AdS$, this singularity structure  can play an important role for light glueball production, and this has been addressed  in \cite{Hong:2005np}. In particular, with $\kappa^{-1}$ large, each of the two terms leads to separate ``diffusion" effects. However, for our present purpose, this  turns out to be less important.   Indeed, from flat-space string theory, one finds that, in the limit $\kappa = \rightarrow \infty$
\begin{equation}
{\cal V}_{flat}(t_1, t_2,\kappa)\rightarrow  \kappa^{-2} \beta_0(t_1) \beta_0(t_2)
\label{eq:largekappa}
\end{equation}
thus leading to a  simplified structure. Furthermore,  since $\kappa = q^2_\perp  + m^2_n$  for high-mass  $\kappa>> q^2_{\perp}$, the dependence on $q_\perp^2$  is effectively lost   up to  $O(q^2_{\perp}/m^2_n)$ corrections.  In moving to impact space, we find that the vertex depends only on $x_2=|x_{2 \perp}-x_\perp|$ and $x_4=|x_{4 \perp}-x_\perp|$, and is independent of $x_{24}=|x_{2 \perp}-x_{4\perp}|$. That is, for production of a heavy object, angular correlation between the left- and right-moving systems decouple, leading to a much simplified analytic structure.  In what follows, for Higgs production, due to $\Lambda_{QCD}<<m_H $, we drop the dependence on $\kappa$ entirely.

With these considerations, we are led to a parametrization of the amplitude for a diffractive high-mass glueball production 
\begin{equation}
A(s,s_1,s_2, t_1,t_2)= \Phi_{13}*\widetilde{\cal K}_P *V_n*\widetilde{\cal K}_P* \Phi_{24} \; .
\end{equation}
where the production vertex, $V_n$, is local in $AdS$-radius, i.e.,
 \bea
A(s,s_1,s_2, t_1,t_2) &\simeq &  \int dz_1 dz  dz_2\; \sqrt{-g_1}\sqrt{-g} \sqrt{-g_2}\;\Phi_{13} (t_1,z_1) \;\widetilde{\cal K}_P(s_1,t_1,z_1,z)      \nn
&&\times    V_n( t_1,t_2,z) \; \widetilde{\cal K}_P(s_2,t_2,z,z_2) \;\Phi_{24}(t_2,z_2)  \;.
\eea

\subsection{Higgs Production and Bulk to Boundary Propagator} 
\label{sec:BtoB}

Returning to the original problem, we now consider the coupling for the Pomeron fusion
to an external source $F^2$ through the  bulk to boundary propagator $K(x_\perp - x'_\perp,z)$.
That is, we are  calculating an amplitude involving a nonnormalizable wave function for the leg associated with the Higgs. Therefore, the amplitude involves one factor of the bulk-to-boundary propagator  $K(q^2,z)$, in a momentum representation.
Using $\Lambda_{QCD} \ll m_H$ we
will show that the dominant contribution to the resulting integration over the AdS radius  comes from
\begin{equation}
z \simeq O(1/M_H).
\end{equation}
To see how this comes about, we need to evaluate the bulk to boundary propagator at $q^2 =- m^2_H$. This issue was studied in some detail by Hong, Yoon and
Strassler~\cite{Hong:2004sa} focused on the vector current coupling
to the tower of AdS glueballs. Here we are considering the scalar term
in the energy-momentum current which is essentially a trivial
modification. More difficult is understanding the analytic continuation to
time-like momentum $q^2 = - m^2_H$.

Consider the hard wall model for a
confining AdS/CFT for clarity at first. Following
Ref. \cite{Hong:2004sa} the bulk to boundary propagator can be viewed
as the non-normalizable wave function excited by the current ($F^2$
in our case) at the boundary:
\begin{equation}
\psi(q^2, z) = \sum_n \frac{f_n \phi_n(z)}{q^2 + m^2_n}
\label{eq:discrete}
\end{equation}
where the ``decay constant" for the n-th glueball state, $f_n$, is
\begin{equation}
f_n = \<0 | F^2(x=0) |n,p\> \sim m^3_n
\end{equation}
for the nth scalar glueball with on-shell eigen solution, $\exp[i p x]
\phi_n(z)$ for $p^2 = - m^2_n$.  Since the bulk-to-boundary propagator is determined by the background geometry, the relative strengths of $f_n$ are fixed.

In the conformal limit, scalar bulk-to-boundary propagator in a momentum presentation can be expressed explicitly is
\begin{equation}
K (z,q) = (q z)^2 K_2(qz) = z^2 \int^\infty_0 \frac{dm m^3 J_2( m z)}{q^2 + m^2} =
\int^\infty_0\frac{dx x^3 J_2( x)}{z^2 q^2 + x^2} \; .
\end{equation}
where $K_\nu$ is the modified Bessel function. Note $(q z/2)^\nu   K_\nu(qz) \simeq \Gamma(\nu)/2 $ for small $qz$  and
$K_\nu(qz) \simeq \sqrt{\pi/(qz)} e^{ - qz}$ for  large $qz > 0$. It follows that, when averaged  over smooth functions of z, integrals such as
\begin{equation}
I(q) = \int_0^\infty dz\;  T(z)\; K(z,q) \sim T(1/q)
\label{eq:endpoint}
\end{equation}
are  dominated by the end-point region where $z=O(1/q)$ as was
also the case for DIS (\cite{Brower:2010wf}).

In our current application to Higgs production, we also need to analytically continue to the region where $q$ is time-like. Numerically we find this is a good approximation after analytical continuation to $- q^2 = (m_H + i \Gamma)^2$ or $q = i m_H + \Gamma$ . We note that we need to convolute this bulk-to-boundary propagator over products of two Pomeron Green's functions, which will be smooth in z. For example  at $t_1 = t_2=0$, the explicit expression, up to smooth logarithmic corrections, can be expressed as
\bea
V(m_H)  &\sim& \pi (z_1z_2)^2  \int^\infty_0 \frac{dz }{z}  e^{\textstyle -  i z  m_H - z \Gamma }    \nonumber \\
&&
 e^{\textstyle  j_0( \tau_1+ \tau_2)
  - \ln^2(z/z_1)/(2- j_0)\tau_1
  - \ln^2(z/z_2)/(2- j_0)\tau_2}
\eea
with $\tau_i = \ln(z z_i s_i)$.  Note the dominant rapid oscillations due to the factor $e^{\textstyle -  i z  M_H}$ and also a convergence factor $e^{\textstyle  - z \Gamma }$. (For confining situation, e.g., hard-wall, the $z$-integral is cutoff at large $z$.)  It can be shown that the  dominant contribution to the
integral for large $M_H$ comes again from $z \sim 1/m_H$. This can be seen by writing the integral as
\begin{equation}
(z_1 z_2)^2  \int^\infty_0 dz \exp[- F(z)]
\end{equation}
where
\bea
F(z) &=&   i z m_H + z \Gamma   - (2 j_0 -1 )\ln(z)  \nonumber \\ 
&+& ln^2(z/z_1)/((j_0 -2)\ln(zz_1 s_1)) + ln^2(z/z_2)/((j_0 -2)\ln( z z_2 s_2))  \; .
\label{eq:F}
\eea
For $s_1 = s_2 = O(\sqrt{s})$ large, the saddle point $z^*$ is essentially determined  by the first two terms on the right of (\ref{eq:F}), leading to
\begin{equation}
z^*\simeq (2j_0-1) /( i  m_H +  \Gamma)
\end{equation}
The validity of this approximation can also be verified numerically.  Therefore as an approximate treatment, one can  represent the bulk-to-boundary propagator by
\begin{equation}
K(x'_H - x_H,z') \sim  \; \delta(x'_H-x_H)\; \delta(m_H z' -c )
\label{eq:saddleHiggs}
\end{equation}
where $c=O(1)$.  We can further rescale $z$ so that $c=1$.

We are finally in the position to put all the pieces together. Although we eventually want to go to a coordinate representation in order to perform eikonal unitarization, certain simplification can be achieved more easily in working with the  momentum representation. The Higgs production amplitude, schematically given by (\ref{eq:adsDoublePomeronScheme}), can be written explicitly as
 \bea
A(s,s_1,s_2, t_1,t_2)&\simeq &  \int dz_1 dz  dz_2\; \sqrt{-g_1}\sqrt{-g} \sqrt{-g_2}\;\Phi_{13} (z_1)     \nn
&\times&   \widetilde {\cal K}_P(s_1,t_1,z_1,z)  \; V_H(q^2 , z)\; \widetilde {\cal K}_P(s_2,t_2,z,z_2)\; \Phi_{24}(z_2)  \;.
\label{eq:adsDoublePomeronHiggs}
\eea
where  $q^2= -m_H^2$.   For the production vertex, however, we will keep  it simple by  expressing it more generally as
\begin{equation}
 V_H( q^2,z)= V_{PP\phi} K(q^2,z) L_H
  \label{eq:HiggsVertex}
\end{equation}
where $K(q^2,z)$ is the conventionally normalized bulk to boundary propagator, $V_{PP\phi}$ serves as an overall coupling from two-Pomeron to $F^2$, and $L_H$ is the conversion factor from $F^2$ to Higgs. Treating the central vertex $ V_{PP\phi}$ as a constant, which follows from the super-gravity limit, we have ignored possible additional dependence on $\kappa$, as well as that on $t_1$ and $t_2$. This approximation gives an explicit factorizable form for Higgs production.

\newpage
\section{Strategy for Phenomenological Estimates}
\label{sec:strategy}

While the goal of this paper is simply to extend holographic diffraction to Higgs production, it is useful to outline the phenomenological approach we plan to pursue to confront experimental data. There should be a strong warning however that details will necessarily change as we discover which parametrization are critical to a global analysis of data.
 Our current version for the holographic Higgs amplitude involves 3  parameters: (1) the hardwall IR cut-off determined by the glueball mass, (2) the leading singularity in the $J$-plane determined~\footnote{Note that in a true dual to QCD at $N_c = \infty$ the only free
parameter is a single mass scale that by ``dimensional transmutation'',   replaces the coupling constant and all
dimensionless ratios. So only the first parameter would be needed phenomenologically.} by the 't Hooft parameter $g^2N_c$ and (3) the strength of the central vertex parametrized by the string coupling or Planck mass. A strategy must be provided for fixing  these parameters.  The glueball mass at large $N_c$ can in principle be  fixed by lattice QCD methods and the leading 
$J$-plane singularity from total cross section data.  Finally the central vertex, $V_H$, or equivalently,  $ V_{PP\phi}$,  via (\ref{eq:HiggsVertex}),  can be normalized, following  the approach of Kharzeev and Levin~\cite{Kharzeev:2000jwa} based on the analysis of the trace anomaly. Since one 
can in principle use  elastic scattering to normalize the bare BPST Pomeron coupling to external protons, once $ V_{PP\phi}$ is known, the central Higgs production amplitude, (\ref{eq:HiggsVertex}), is completely determined. This procedure for determining this crucial overall scale deserves further explanation.

\subsection{Continuation to Tensor Glueball Pole and On-Shell Higgs Coupling}

Confinement deformation in AdS will lead to glueball states, e.g., the lowest tensor glueball state lying on the leading Pomeron trajectory~\cite{Brower:2000rp}.   There will also be scalar glueballs associated with the dilaton.  With scalar invariance broken, this  will also lead to non-vanishing couplings between a pair of tensor glueballs and scalar glueballs. When translated into a Witten diagrammatic description in the bulk, as we have also pointed out earlier,
this leads to a non-vanishing graviton-graviton-dilaton coupling and  in turn leads to  $V_H\neq 0$.

Consider first the elastic amplitude. With confinement,
each Pomeron kernel  will contain a tensor glueball pole when $t$ goes on-shell. Indeed, the propagator for our Pomeron kernel can be expressed as a discrete sum over pole contributions, i.e., Eq. (\ref{eq:spectrumqIR}).   That is, when $t\simeq  m_0^2$, where $m_0$ is the mass of the lightest tensor glueball, which lies on the leading Pomeron trajectory, i.e., $m_0^2= m_0^2(j=2)$, we need only to keep the lowest term in  (\ref{eq:spectrumqIR}), and the kernel takes on a factorizable form.

 It follows that the elastic amplitude, Eq. (\ref{eq:adsPomeron}),   when approaching the $t$-channel pole at $t= m_0^2$, is  pole-dominated,
\begin{equation}
A(s,t) \simeq g_{13}\; \frac {s^2}{m^2_0-t} \; g_{24}
\label{eq:2to2KL}
\end{equation}
with vertex $g_{ij}$  given by an overlapping integral:
\begin{equation}
g_{ij}(m_0^2)=\alpha'   \int dz \sqrt{-g(z)} e^{-4A(z)}  \;\beta(m_0^2) \Phi_i(z)\Phi_j(z) \; \phi_{G}(z) \; .
\label{eq:externalcoupling}
\end{equation}
The tensor glueball wave function, for hard-wall model, is given by
$\phi_{G}(z) = \widetilde \phi_0(x,j=2)$. 
Here, we have generalized   $\Phi_{ij}$  as  the product of two external wave functions, together with an extra factor $\beta(t)$, i.e., $\Phi_{ij} (t,z)=\beta(t) e^{-2A(z)}\Phi_i(z)\Phi_j(z)$, for phenomenological  reasons.  That is, the external coupling  $g_{ij}$ is given by an overlap-integral over a product of three wave functions,   $\Phi_i(z)$, $ \Phi_j(z)$  and $  \phi_{G}(z)$, consistent with the discussion in~\cite{Hong:2005np,Hong:2004sa}. With the standard normalization, 
 $A(s,t)$ is dimensionless, thus $g_{ij}(m_0^2)$ has the dimension of length.

Let us next turn to  the Higgs production amplitude, Eq. (\ref{eq:adsDoublePomeronHiggs}). Note that the Pomeron kernel now appears twice, $\widetilde {\cal K}_P(s_1,t_1,z_1,z) $ and $\widetilde{\cal K}_P(s_2,t_2,z_2,z) $.   When nearing the respective tensor poles at $t_1\simeq  m_0^2$ and $ t_2\simeq  m_0^2$,  we can again use the leading-pole approximation for  the kernel, (\ref{eq:leadingpole}).  The amplitude can now be expressed as
\bea
A(s,s_1,s_2, t_1,t_2)&\simeq&g_{13}\; \frac{  \Gamma_{GGH} \;  s^2  }{(t_1-m_0^2)(t_2-m_0^2)  }   \; g_{24}
   \label{eq:2to3KL}
\eea
We have performed the $z_1$ and $z_2$ integrations as done for the elastic amplitude,  leading to external coupling $g_{13}$ and $g_{24}$ respectively, given by (\ref{eq:externalcoupling}), and  also have made use of the fact that $s_1s_2\simeq \kappa\; s\simeq m_H^2 s $. 
Here $\Gamma_{GGH}$ is the effective on-shell glueball-glueball-Higgs coupling, which can be written in terms of the central vertex,  $V_{PP\phi}$.

Recall that, 
after integrating out the
top quark, the effective on-shell glueball-glueball-Higgs coupling can be expressed as
$ \Gamma_{GGH} =L_H  F ( - m^2_H)$,
with $L_H$ given by (\ref{eq:LH})  and    
\bea
F(q^2) &=&(\alpha' m_H^2)^2   V_{PP\phi} \int dz \sqrt{-g(z)} e^{-4A(z)} \phi_G(z) K(q,z) \phi_G(z)
\label{eq:formfactor}
\eea
This follows from (\ref{eq:adsDoublePomeronHiggs}) and (\ref{eq:HiggsVertex}).  Observe that, by going one shell,  $ F $  can be considered as a scalar form factor, where
\begin{equation}
F(q^2) = \<G, ++,q_1| F^a_{\mu\nu} F^a_{\mu\nu}(0) | G,--,q_2\>
\end{equation}
That is, in the high energy Regge limit, the dominant contribution comes from  
 the maximum helicity glueball state~\cite{Brower:2006ea}, with $\lambda = 2$.   Note that (\ref{eq:formfactor}) is in the form of an integral over a product of  the bulk-boundary propagator with other smooth functions, (\ref{eq:endpoint}). In the large $m_H$ limit, the dominant contribution comes from $z=O(m_H^{-1})$.  What remains to be specified is the overall normalization, $F(0)$, or equivalently the coupling $V_{PP\phi}$.

We also note in passing that the expression for this form factor  is precisely what is expected from Eq. (\ref{eq:WittenVertex}) since we are considering a situation with a constant Witten vertex, assuming no running~\footnote{A more consistent treatment would be to adopt the precise coupling given by  the model, (\ref{eq:WittenVertex}). However, we will not follow this path in what follows.}.  Conformal scaling determines the large $Q^2=|q^2|$ behavior
to be
$
F_{GG}(q^2)  \sim (1/Q)^{\Delta_1 + \Delta_2 - 4} = 1/Q^4
$
with $\Delta_1 = \Delta_2  = \Delta_G= 4$,  as can be readily checked.
This form factor can also be written, using (\ref{eq:discrete}),  as
\be
F_{GG}(q^2) = (\alpha' q^2)^2 \sum_n \frac{ g_{GG\varphi_n}\; f_n}{q^2 + m^2_n}\; ,
\ee
which follows when we express the bulk-to-boundary propagator $K(q,z)$  in a dispersion sum  over
the dilaton scalars,  with  the coupling constants given by
\begin{equation}
g_{GG\varphi_n} =   V_{PP\phi} \int dz  \sqrt{-g(z)} e^{-4A(z)} \;  \phi_G(z) \phi_n(z)  \phi_G(z)
\end{equation}

\subsection{Normalization of the Glueball Form Factor}
\label{sec:anomaly}

From the gauge theory view, we recognize that $F(q^2)$ at $q^2=0$, is a matrix element of the trace of the energy momentum tensor.
So we  can follow  D. Kharzeev and E. M. Levin  \cite{Kharzeev:2000jwa}, who  considered the matrix elements of  the trace-anomaly
between two states, $|\alpha(p)\>$ and $|\alpha'(p')\>$, with four-momentum transfer $q=p-p'$. In particular, for a single particle state of a tensor glueball $|G(p)\>$, this leads to $\<G(p)|  {\Theta}^\alpha_\alpha|G(p')\> = \frac{\widetilde \beta}{2 g} \<G(p)| F^a_{\mu \nu} F^{a \mu \nu} | G(p')  \>$. 
At $q = 0$, the forward matrix element of the trace of the energy-momentum tensor is given simply by  the mass of the relevant tensor glueball, with  $\<G|{\Theta}^\alpha_\alpha | G  \> = M_G^2$, this directly yields
\begin{equation}
F(0) = \<G| F^a_{\mu \nu} F^{a \mu \nu} | G  \>  =-  \frac{4\pi  M_G^2}{3 \widetilde \beta }
\label{eq:FF}
\end{equation}
where $\widetilde \beta = - b\alpha_s/(2 \pi)$, $b = 11 - 2n_f/3$, for $N_c=3$.  In what follows, we will use $n_f=3$.   Note that heavy quark contribution is not included in this  limit.  Since the conformal scale breaking is due to the running coupling constant in QCD, 
there is apparently a mapping between QCD scale breaking and breaking of
the AdS background  in the IR, which  gives a finite mass to the glueball
and to give a non-zero contribution to the gauge condensate.

We are now in the position to provide the proper normalization for $ V_{PP\phi}$. It is convenient to formally define
\begin{equation}
 \Gamma_{GGH}(q^2) \equiv  \frac{\alpha_s g}{24 \pi M_W} F( q^2)
\end{equation}
It follows that  the normalization  at $q^2=0$ is given by
\begin{equation}
\Gamma_{GGH} (q^2=0)= \frac{\alpha_s g}{24 \pi M_W} F(0)=  \frac{ 2M^2_G}{3 v b} = 2^{1/4}
G^{1/2}_F \frac{ 2M^2_G}{27}
\label{eq:anomaly}
\end{equation}
More explicitly, from (\ref{eq:adsDoublePomeronHiggs}) and (\ref{eq:HiggsVertex}), we have   
\bea
\;V_{PP\phi}  =C^{-1}  L_H^{-1} (\alpha' m_H^2)^{-2 } \Gamma_{GGH}(0)
\label{eq:centralvertex}
\eea
where $C$ is given by a finite integral
\begin{equation}
C=\int dz \sqrt{-g(z)} e^{-4A(z)} \phi_G(z) K(0,z) \phi_G(z)
\label{eq:C}
\end{equation}
 and is $O(1)$.

 It is worth recapitulating the key discussion  in  \cite{Kharzeev:2000jwa}, which makes essential use of the  scale anomaly. More generally, when  quarks are massive, trace anomaly can be expressed as 
  \begin{equation}
{\Theta}^\alpha_\alpha= \frac{\widetilde \beta}{2 g} F^{\alpha\beta a}F_{\alpha\beta a} + \sum_{l=u,d,s} m_l \bar q_l q_l +  \sum_{h=c,b,t} m_h \bar q_h q_h
\label{eq:traceAnomalyfull}
\end{equation}
where we have separated quarks into light and heavy. In (\ref{eq:traceAnomalyfull}), $\widetilde \beta$ now includes contributions from all quarks, both light and heavy. In this case, scale invariance is broken explicitly by quark masses. In the heavy quark decoupling limit, it has been shown that the explicit contribution from the last term cancels  the corresponding heavy quark contribution to  the  beta-function.   For simplicity, consider also the situation where light quarks are massless.  In this limit, $\widetilde \beta$ receives contributions from gluons and light quarks only. Therefore, a proper heavy quark decoupling in (\ref{eq:traceAnomalyfull}) allows an identification of   matrix element of $m_h \bar q_h q_h$ in the heavy quark limit with that of the corresponding matrix element of $F^2$. This equivalence corresponds precisely to the mechanism 
 for a central Higgs production discussed in Sec.~\ref{sec:PPfusion}.
 
Recall that  Higgs couples directly  to quark-antiquark pairs, Eq. (\ref{eq:HiggsCoupling2ttbar}), with a strength proportional to the quark mass.  In practice, we will keep only the $t \bar t$ pair.  In our earlier treatment, after taking into account the triangle graph, it converts this coupling to an effective local interaction directly with gluons through the combined effective Higgs production coupling where Higgs now couples directly to $F^2$, going from (\ref{eq:HiggsCoupling2ttbar}) to (\ref{eq:effectiveHiggsCoupling}).
This conversion is achieved in the limit  $|q|/m_t\rightarrow 0$ with $M_W$ related to the vev~\footnote{Instead of $v^{-1}$, this can also be expressed  in term of the Fermi coupling, $G_F$, where  $\sqrt{2} G_F \equiv (g/2 M_W)^2 = (1/v)^2$ .} by $2 M_W = g v$.  This limit corresponds to the ``decoupling limit" where heavy quarks contribution to the QCD beta-function disappears, consistent with the discussion above for decoupling in the trace relation,  (\ref{eq:traceAnomalyfull}).
The effective on-shell glueball-glueball-Higgs coupling, $ \Gamma_{GGH} $,  can now be obtained in this limit from  (\ref{eq:effectiveHiggsCoupling}) by taking the matrix element, $ \Gamma_{GGH}  \simeq  \< 0|\int {\cal L}|\phi_H GG\>$.  This corresponds to an effective replacement of $\< G|t\bar t |G\>$  by  $\< G| F^2 |G\>$.  To be more precise,  we have
\bea
 \Gamma_{GGH}(q^2) \simeq   \frac{ g m_t}{2 M_W} \<G|   t \bar t |G\>   &\simeq & \frac{\alpha_s g}{24 \pi M_W} \<G|  F^a_{\mu\nu}F^{a  \mu \nu}(0) |G\>\nn
 & =& \frac{\alpha_s g}{24 \pi M_W} F_{GG} (q^2)
\label{eq:GGHiggs}
\eea
with $q^2=-m_H^2$. Eq.  (\ref{eq:anomaly}) serves to provide a normalization and continuation to physical $q^2$ can be done through (\ref{eq:formfactor}). With $m_H^2$ large, this integral can be estimated by the ultra-local approximation discussed in Sec. (\ref{sec:BtoB}). Observe that $\Gamma_{GGH}$ has the dimension $O(E)$, as expected. Even more importantly, the only relevant mass scales are $M^2_G$ and the vev;  the mass of Higgs does not enter.  For more discussion on this point, see \cite{Kharzeev:2000jwa}.

Since  (\ref{eq:FF}) plays a key role for our estimate  for diffractive Higgs production, we provide below a more direct  but equivalent derivation using Feynman-Hellman Theorem, without explicitly invoking trace-anomaly. All dimensionful quantities can be written in terms of $\Lambda_{QCD}$, or equivalently the glueball mass, 
\begin{equation}
M_G(\alpha_s(\mu),\mu) = c_o \Lambda_{QCD} =
c_o \mu e^{\textstyle - \int^{\alpha_s(\mu)}\frac{d\alpha_s}{\alpha_s \widetilde \beta} }
\end{equation}
where~\footnote{The standard definition of the beta function give
$\frac{\partial g}{\partial \mu} = \frac{g^3}{16 \pi^2}[11 N_c/3 - 2 n_f/3] \rightarrow  \frac{g^3}{16 \pi^2}b$, with $b = 9$ for $N_c = 3, n_f = 3$. Here we use $\widetilde \beta(g) = 2 \beta(g)/g$.  }
\begin{equation}
\frac{\mu}{\alpha_s} \frac{\partial\alpha_s}{\partial \mu} = 2 \beta(\alpha_s)/g_s \equiv \widetilde \beta(g)
\end{equation}
Since masses  must be independent of the choice of
renormalization scheme, one has
\begin{equation}
0= \mu \frac{d}{d\mu} M_G =  M_G -  \frac{\partial M_G}{\partial \alpha_s}  \mu
\frac{\partial\alpha_s}{\partial \mu} = M_G-  \frac{M_G}{\widetilde \beta(\alpha_s) \alpha_s} \mu
\frac{\partial\alpha_s}{\partial \mu}
\end{equation}
This is of course equivalent to
\begin{equation}
\alpha_s\frac{\partial}{\partial \alpha_s} M_G =  -\frac{M_G}{ \widetilde \beta(\alpha_s) }\, .
\end{equation}
 Now we also can apply the Feynman-Hellman theorem to compute the
derivative 
\begin{equation}
1/\widetilde \beta(\alpha_s) = - \alpha_s \frac{\partial \log M_G}{\partial \alpha_s}
= - \frac{1}{16 \pi \alpha_s} \frac{\<G| G^a_{\mu\nu}G^{a \mu \nu}(0)
|G\>}{M_G} \times \frac{V_3}{\<G|G\>}
\end{equation}
where the  Hamiltonian has been taken in the form: $H = - (1/16 \pi \alpha_s) \int_{V_3} d^3x G^a_{\mu\nu}G^{a \mu \nu} + \cdots$.  Taking
the infinite volume limit $V_3 \rightarrow \infty$ and  going back to the conventional normalization for the gauge field, one arrives at 
\begin{equation}
\<G| F^a_{\mu\nu}F^{a \mu \nu}(0)
|G\>  = - 4 \pi M^2_G/ 3\widetilde \beta(\alpha_s)
\end{equation}
as expected.

\subsection{Near-Forward limit}

With the Higgs production vertex  fixed by (\ref{eq:centralvertex}), all factors for the Higgs production amplitude, (\ref{eq:HiggsVertex}), are in principle determined, and we are now in the position to turn to the physical region where $t_1\leq 0$ and $t_2\leq 0$. To complete the discussion, let us examine here more closely the Pomeron kernel, leaving the phenomenological  question of elastic vertex to the next section.  

 Consider first  the elastic amplitude.  In a strict supergravity limit,  the kernel follows from a simple pole-dominance ansatz, Eq. (\ref{eq:2to2KL}), which   is clearly inadequate at $t  = 0$ since the amplitude remains real. However, with  $\lambda$  finite, 
 as one moves from $t\simeq m_0^2$ to $t\simeq 0$, the amplitude becomes complex, with the leading $s$-dependence slowing down from $s^2$ to an non-integral power.  To carry out this analysis, as we have explained earlier,  it is necessary to revert to the $J$-plane representation, Eq. (\ref{eq:PomeronKernel}), for the Pomeron kernel $\tilde {\cal K}_P(s,t,z,z')$. When confinement deformation is implemented,  the $J$-plane propagator  $\widetilde G_j(t,z,z')$ can be expressed generally in a spectral representation, (\ref{eq:spectrumqIRGeneral}). Equivalently, it can also be expressed as  a sum of $J$-dependent poles in $t$, e.g., for hardwall model, 
 \bea
\widetilde G_j(z,z';t)
&=&  \sum_n \frac{ \widetilde \phi_n(z,j)\widetilde \phi_n(z',j) }{m^2_n(j) -t} \;.
\label{eq:spectrumqIR}
\eea
with $\widetilde \phi_n(z,j)$ given in terms of Bessel functions. 
 As one moves away from region near the tensor pole, the leading $J$-plane structure initially remains a Regge pole,   As we move close to the $t=0$ boundary, the $J$-plane structure  is highly model-dependent. If asymptotic freedom is implemented, the amplitude remains meromorphic in $J$, and  a leading pole approximation can remain meaningful.  Realistically, due to diminishing trajectory spacings, a superposition of a large number of poles will be required.  (See Appendix-A for more details.) 

In what follows, we shall use the hardwall model as a guide. One finds the amplitude is dominated by a Regge pole initially in a region $m_1^2<t<m_0^2$, where $m_1^2$ is the point where the leading pole disappears through the BPST cut at $j_0=2-2/\sqrt\lambda$, i.e., $m_0(j_0)=m_1$. As one further continues to the physical region where $t\leq 0$, the amplitude will now be dominated by the contribution from the BPST cut, with  the inverse Mellin transform in $J$ turning into an integral over the discontinuity across the cut.
 This  leads  to  an explicit AdS representation for the Pomeron kernel in the near-forward limit, 
\begin{equation}
\widetilde {\cal K}_P(s,t,z,z') =   \int_{-\infty}^{j_0} \frac{dj}{2\pi i} \xi(j) (
\alpha'  \widehat s)^{j} \; {\rm Disc}_j\;    \widetilde G_j(t,z,z') \; .
\label{eq:PomeronKernelcut}
\end{equation}
Given the Pomeron kernel, both the elastic amplitude (\ref{eq:adsPomeron}) and the diffractive central Higgs production amplitude,  (\ref{eq:adsDoublePomeronHiggs}), are now completely specified, as promised.

For completeness, we note that analytic expression for the Pomeron kernel for the  hardwall model can be found in \cite{Brower:2006ea,Brower:2010wf}, and it takes on a relatively simple form in the forward limit, i.e., at  $t=0$. For illustrative purpose, it is useful to exhibit its form in the conformal limit in an impact-representation. In this representation, one finds, for the imaginary part of the kernel~\cite{Brower:2007xg},
\be
{\rm Im}\;\; \widetilde{\cal K}_P= \frac{1}{\pi}\frac{1}{z z'}\, \frac{\xi}{\sinh\xi}\, (\sqrt{\lambda}/32\pi)^{1/2}\, e^{j_0\, \tau}\, \frac{e^{-\sqrt{\lambda}\xi^2/2\tau}}{\tau^{3/2}}\, ,
\label{eq:conformalkernel}
\ee
where $\tau = \log(\alpha'\widehat{s}),\, e^{\xi} = 1+v+\sqrt{v(2+v)},$ and $v$ is the $AdS_3$ chordal distance squared, $v = ({\bf x_\perp}^2 + (z - z')^2)/2zz'.$   While the above equation is given in impact parameter space, we can perform a Fourier transform to go to momentum space. This has a particularly simple form at $t = 0$, where the $b$-space integral can be performed explicitly to obtain
\begin{equation}
{\rm Im}\;\; \widetilde{\cal K}_P(s,0,z,z') =  \left(\frac{1}{8\pi\sqrt{\lambda}}\right)^{1/2}\, e^{j_0 \tau}\, \frac{\exp{\left[-\frac{\sqrt{\lambda}}{2\tau}(\log z - \log z')^2\right]}}{\tau^{1/2}}\, ,
\label{eq:conformalkerneltzero}
\end{equation}
which corresponds  to a diffusion kernel in the AdS-radius.   The imaginary part of the Pomeron kernel in the hardwall model similarly has a simple closed form expression at $t = 0$ 
\begin{equation}
{\rm Im}\;\; \widetilde{\cal K}_{HW}(s,0,z,z') =  \left(\frac{1}{8\pi\sqrt{\lambda}}\right)^{1/2} e^{j_0 \tau} \left(\frac{e^{-\frac{\sqrt{\lambda}}{2\tau}\log^2(z/z')}}{\tau^{1/2}}\,
+F(z,z,\tau)\frac{e^{-\frac{\sqrt{\lambda}}{2\tau}\log^2(zz'/z_0^2)}}{\tau^{1/2}}\right) . 
\label{eq:hardwalltzero}
\end{equation}
The function 
\begin{equation}
  F(z,z',\tau) = 1-2\sqrt{\rho\pi\tau}e^{\eta^2}\, erfc(\eta)\, , \,\,\,\, \eta = \frac{-\log{z z'/z_0^2} + 2\tau/\sqrt{\lambda}}{\sqrt{2\tau/\sqrt{\lambda}}}\, ,
\end{equation}
is fixed by the boundary conditions at $z_0$, and it displays the relative strength of confinement. For a discussion, see \cite{Brower:2006ea,Brower:2010wf}. The real part of the kernel does not have a simple closed form expression, however it was shown in \cite{Brower:2007xg} that in the limit of large $\hat{s}$ with $\lambda$ large and fixed, so that $\sqrt{\lambda}/  \log(\alpha'\widehat{s}) \ll 1$, the relationship between the real and imaginary parts  takes on a simple form
\begin{equation}
  {\rm Re}\;\; \widetilde {\cal K}_P = \cot(\frac{\pi}{\sqrt{\lambda}})\, {\rm Im}\;\; \widetilde {\cal K}_P\, .
\end{equation}


\subsection{First Estimate for Double-Pomeron Contribution}
 \label{sec:guess}
 
We now turn to a  qualitative discusion for estimating the convolution over the AdS/QCD building blocks that determine Higgs production in our model. We emphasize
however that in fact all the kernels are defined precisely through the differential equations and therefore with sufficient numerical effort can be computed directly once the explicit model AdS background is chosen and the proton impact factor is fixed.  This  numerical step is premature and is postponed to a  future phenomenological analysis.

Even to provide a  rough phenomenological estimate for the central Higgs production cross section, we cannot avoid dealing with the coupling of Pomeron to the external protons. At present we are using a 
very naive model of the AdS proton. Below we treat the proton as a glueball wavefunction modified by a form factor for non-zero $t$. It is worth noting that an even
simpler wave function, treating the proton as  a ``heavy'' point like object at the IR cut-off, has already  provided surprisingly 
good  fits  to the HERA DIS and DVCS data at small x. We anticipate the need to improve this in self-consistent fits. Indeed modeling the proton in the AdS context is an interesting topic in itself~\cite{Hashimoto:2010je,Hashimoto:2008zw,Yi:2008zz,Domokos:2010ma,Domokos:2009hm}, as mentioned in the Introduction.   As in the case of elastic scattering, it is also pedagogically reasonable to begin by first treating the simplest case  of double-Pomeron exchange for Higgs production, i.e., without absorptive correction. 
Here, we discuss how phenomenologically reasonable simplifications can be made. This is followed by treating eikonal
corrections in the next section, which provides a means of  estimating the  all-important survival probability. 

It is often useful to make use of the conformal kernel at $t=0$ \cite{Brower:2006ea,Brower:2007xg} where diffusion in the AdS radius is evident.  The confinement can also be treated at $t=0$, leading to (\ref{eq:hardwalltzero}).  The  structure of the  kernel at $t<0$, can in principle be obtained by solving the appropriate DE for the $j$-plane propagator. It can also be treated approximately, e.g., iteratively in an expansion about the known  solution at $t=0$.  Alternatively, it is more useful to assume  pole-dominance, e.g.,  keeping  the  contribution coming  from a leading trajectory, even for $t$ negative, 
\begin{equation}
\widetilde G_j(z,z';t)\simeq  \widetilde \phi_{eff}(z,j)   \frac{  1 }{m^2_{eff}(j) -t}   \widetilde \phi_{eff}(z',j) \;.
\label{eq:leadingpole}
\end{equation}
where we approximate the  BPST-cut contribution by that of an effective leading pole,  with  the Pomeron kernel behaving as $s^{j_{eff}(t)}$, where  $j_{eff}(t)$, the trajectory function, determined by $m^2_{eff}(j(t)) = t$. That is, we assume $j_{eff}(t)$  remains real in the physical region where $t<0$. By performing the inverse Mellin transform, (\ref{eq:PomeronKernel}), the large $s$-behavior of the BPST kernel 
can easily be obtained, leading to 
\begin{equation}
A(s,t) \simeq g_{13}(t)\;\left( \frac{ \xi(j_{eff}(t))\; ( \alpha' s)^{j_{eff}(t)}}{\alpha'^{2}\widetilde m^2(t)  }\right)  \; g_{24}(t)
\label{eq:2to2P}
\end{equation}
where   $\widetilde m^2$ is the inverse of trajectory slope, $\widetilde m^2(t) \equiv  dm^2_{eff}(j(t))/dj$, and  $\xi(j)$ is the signature factor.
For the elastic amplitude, the coupling
\begin{equation}
g_{ij}(t)= \alpha'  \int dz \sqrt{-g(z)}  \;\Phi_{ij}(t,z) e^{-j_{eff}(t)A(z)} \widetilde \phi_{0}(z,j_{eff}(t))
\end{equation}
is again in the form of an overlapping integral over the product of three wave functions, with $ \widetilde \phi_{eff}=\widetilde \phi_{0}(z,j_{eff}(t))$.  This serves as a continuation away from the on-shell spin-2 exchange by  replacing the spin-2 wave function $\phi_G(z)= \widetilde \phi_0(z,2)$ by a corresponding wave function for a Pomeron, $ \widetilde \phi_{0}(z,j_{eff}(t))$, with spin shifted from 2 to $j_{eff}(t)$.  Although this shift is of the order $O(1/\sqrt \lambda)$, it is important to note that
 $\widetilde \phi_0(z,j(t))\sim z^{\Delta(j(t))-2} $, for $z\rightarrow 0$,  in contrast to  $\widetilde \phi_0(z,2) \sim z^{2}$. Note that  we have continued with the convention where $g_{ij}(t)$ has  the dimension of length. It is also easy to check~\footnote{The tensor pole at $t=m_0^2$ is contained  in the signature factor, and,  in order to match (\ref{eq:2to2P}) with  (\ref{eq:2to2KL}) near $t=m_0^2$, a factor of $2/\pi$ has to be supplied. This can easily be absorbed, e.g.,  by a re-definition for the signature factor. We will not be concerned with such a re-definition for our present purpose.}  that (\ref{eq:2to2P}) reduces to (\ref{eq:2to2KL}) as one approaches the tensor pole, $t\rightarrow m_0^2$.

Focussing next on the forward limit $t=0$, we denote the effective intercept by $\bar j_0$ and  inverse slope by $\widetilde m^2$. Together with the forward coupling $g_{ij}(0)$, 
they will be determined phenomenologically.  We note that $\widetilde m^2$ can be chosen to be  of the order of  the tensor glueball mass, $m_0^2$. For consistency, we also assume that $\bar j_0\simeq j_0$. In the high energy limit, $\xi(\bar j_0)$ provides the phase for the amplitude, with 
\begin{equation}
|\xi|^2 =1 +  \rho^2 = 1+ \left (\frac {{\rm Re} \; A(s,0)}{{\rm Im} \; A(s,0)}\right)^2
\end{equation}

A corresponding  treatment at $t_1\simeq t_2\simeq 0$ for  the Higgs production amplitude, Eq. (\ref{eq:adsDoublePomeronHiggs}), can lead to a similar simplification. 
   It follows,  after a bit of algebra,
   \bea
A(s,s_1,s_2, t_1\simeq 0, t_2\simeq 0)&\simeq &  g_{13}(0)\frac{  \xi(\bar j_0)^2 \Gamma_{PPH}\; ({\alpha' } s)^{\bar j_0}    }{(\alpha' \widetilde m^2)^2} \     \;     g_{24}(0)
\eea
with an effective central vertex, related to $V_{PP\phi}$, (\ref{eq:centralvertex}), by
\bea
\Gamma_{PPH}&\simeq& \frac{\alpha_s g}{24 \pi M_W}  V_{PP\phi} \left({\alpha' m_H^2}\right)^{\bar j_0}C(\bar j_0) 
\eea
where 
\begin{equation}
C(\bar j_0) = \int dz \sqrt{-g} e^{-4A(z)} \widetilde \phi_0(z,\bar j_0) K(-m^2_H,z) \widetilde \phi_0(z,\bar j_0)
\label{eq:C0}
\end{equation}
and we have dropped terms lower order in $O(1/\sqrt\lambda)$. 
We point out  that (\ref{eq:C0})  is finite due to the wave-function normalizability. For hard-wall, it is  logarithmically divergent as $\bar j_0\rightarrow j_0$ which corresponds to  the onset of a Regge cut. In  a proper treatment when the leading singularity is  a cut, this apparent divergence will be absent. In order to avoid complicating the discussion, we proceed with the understanding that $C(\bar j_0)$  is of the order unity.

Let us turn next to the non-forward limit.  We  accept the fact that, in the physical region where $t<0$ and small, the cross sections typically have an exponential form, with a logarithmic slope which is mildly energy-dependent. We therefore approximate all amplitudes in the near forward region where $t<0$ and small,
\bea
A(s,t) &\simeq & e^{B_{eff}(s)\; t/2}\;  A(s,0)
\eea
where $B_{eff}(s)$ is a smoothly slowly increasing function of s, (we expect it to be logarithmic).
We also assume, for $t_1<0$, $t_2<0$ and small, the Higgs production amplitude is also strongly damped so that
\begin{equation}
A(s,s_1,s_2, t_1, t_2)  \simeq    e^{B'_eff(s_1) \; t_1/2}  e^{B'_eff(s_2)\; t_2 /2   }  \; A(s,s_1,s_2, t_1\simeq 0, t_2\simeq 0)
\label{eq:higgs}
\end{equation}
We also assume $B'_{eff}(s)\simeq B_{eff}(s) + {\rm  b}$.   With these, both the elastic, the total pp cross sections and the Higgs production cross section can now be evaluated.  Various cross sections will of course depend on the unknown slope parameter, $B_{eff}$, which can at best be estimated based  on prior experience with diffractive estimates.

The phase space for diffractive Higgs production can be specified by the rapidity of Higgs $y_H$, and two-dimensional transverse momenta $q_{i,\perp}$,  $i=3,4,5$, with $q_{5,\perp}=q_{H,\perp}$, in a frame where the incoming momenta $k_1$ and $k_2$ are longitudinal.   Alternatively, due to momentum conservation, we can use instead $y_H, t_1, t_2, \cos \phi$ as four independent variables where  $t_1\simeq - q_{3,\perp}^2$,  $t_2\simeq - q_{4,\perp}^2$, and $\cos \phi=\hat q_{3,\perp}\cdot \hat q_{4,\perp}$.   However, the amplitude is effectively independent of $\phi$ since its dependence enters through the $\kappa$ variable where $\kappa\simeq m_H^2+ q_{H,\perp}^2= m_H^2 + (q_{2,\perp}+q_{4,\perp})^2$. As  discussed earlier, for Higgs production, we can replace $\kappa$ by $\kappa_{eff} \simeq m_H^2$.

Following the earlier analysis, Sec.~\ref{sec:guess}, it is now possible to provide a first estimate for the double-diffractive Higgs production. It is possible to  adopt an approach advocated by Kharzeev and Levin where the dependence on $B_{eff}$ can be re-expressed in terms of other physical observables. Under our approximation, it is easy to show that  the ratio $ \sigma_{el}/\sigma^2_{total}$ can be expressed as
\begin{equation}
 \frac{\sigma_{el}}{\sigma^2_{total}}=\frac{ 1+\rho^2 }{16\pi B_{eff} (s) }
\end{equation}
where  $\rho={\rm Re} A(s,0)/{\rm Im} A(s,0)$.
Upon squaring the amplitude, $A(s,s_1,s_2, t_1, t_2)$, (\ref{eq:higgs}), the double-differential cross section for Higgs production can now be obtained. 
 After integrating over $t_1$ and $t_2$ and using the fact that, for $m_H^2$ large  $s \simeq  {s_1 s_2}/{m_H^2}$,
one finds
\bea
\frac{d\sigma}{dy_H } &\simeq &(1/\pi) \times C'   \times |   \Gamma_{GGH}(0)/\widetilde m^2|^2  \times \frac{\sigma( s)}{\sigma(m_H^2)} \times R^2_{el}(m_H \sqrt s )
\eea
where $C'=(C(\bar j_0) /C)^{2}$. In this expression above, both $C'$ and $\widetilde m^2$, like  $ m_0^2$,  are model dependent. It is nevertheless interesting to note that, since $\Gamma_{GGH}(0)\sim m_0^2$, the glueball mass scale also drops out, leaving a model-dependent ratio of order unity.    In deriving the result above, we have replaced $B'_{eff}$ by $B_{eff}$ where the difference is unimportant at high energy.  With $m_H$ in the range of $100 GeV$, $R_{el}$ can be taken to be  in the range $0.1$ to $0.2$. For $C'\simeq 1$, $\widetilde m  \simeq m_0$, we find  \be
\frac{d\sigma}{dy_H }\simeq .8\sim 1.2  \;\; {\rm pbarn}.
\ee
This is of the same order as estimated in \cite{Kharzeev:2000jwa}.  However, as also pointed in \cite{Kharzeev:2000jwa}, this should be considered as an over-estimate. The major source of suppression will come from absorptive correction, which can lead to a central production cross section in the  femtobarn  range.

\newpage
\section{Summary and  Discussion}
\label{sec:discusion}

There  is of course already a growing literature with long historical roots
applying both gluon perturbation theory and/or Regge
parametrization to analyze diffractive processes.
By extensive calibration with data, these are converging
on estimates for double diffractive Higgs production~\cite{Kharzeev:2000jwa,Khoze:1997dr,Khoze:2000cy,GayDucati:2011zx,Gastmans:2011zz,Bialas1991540,Coughlin:2009tr,Spira:1995rr,Brodsky:2006wb,Ryskin:2009tk}. While
there are still many areas of controversy in these
models, a general agreement is the need to
combine aspects of hard and soft scattering.  These
approaches can help us to   gain further insight into the strong coupling
approach and to guide future attempts to extract phenomenological
constraints.  The chief advantage of our holographic approach is the ability to unify both
soft (Regge)  and hard (BFKL) diffraction. When supplemented with the confinement deformation 
 in AdS space, e.g., a hard-wall cutoff, our approach not only provides a description for the high energy near-forward scattering, but also allows one to  analytically continue the amplitudes to   tensor glueball pole to normalize the amplitude using the trace anomaly. 
Of course, as emphasized in the Introduction, we have but taken the first step in applying the BPST diffractive model for central Higgs production.  What we have achieved so far  is clearly not the final story, e.g., as for elastic amplitudes, it is well known that  higher order contribution must be taken into account in order to restore the  unitarity constraint. 

There are also other considerations  which we have glossed over in this initial effort.   As stressed in the Introduction, the ``bare-bone" double-Pomeron contribution to  diffractive Higgs production requires three building blocks: the proton impact factors, $\Phi_{ij}$ and  the Pomeron kernel (or Reggeon propagators), $\widetilde{\cal K}_P$ and the Pomeron-Pomeron-Higgs vertex $V_H$.  In this paper, we have treated the impact factors phenomenologically.  
In this concluding section, we focus on discussing how consideration of higher order contributions via an eikonal treatment leads to corrections for the central Higgs production. Following by now established usage, the resulting production cross section can be expressed in terms of a ``survival probability" \cite{PhysRevD.47.101,Gotsman:1993vd,Gotsman:1999xq,Gotsman:2008tr,Gotsman:2011xc,Kaidalov:2001iz,Ryskin:2007qx,Ryskin:2009tk}. In a more traditional usage, this eikonal sum can  also  be referred to as the ``Good-Walker" sum \cite{Good:1960ba}, with contributions due to triple-Pomeron effect as coming from ``enhanced  diagrams". 
For now, we shall ignore these complications and focus on the most important aspect which can be thought of as  ``local saturation" in the Euclidean $AdS_3$.

The importance of eikonal correction can be seen as follows. One of the most important findings from String/Gauge duality is the
fact that, at strong coupling, the single ``Graviton'' (Pomeron)
exchange leads to an elastic amplitude which grows too fast, $\sim
s^{j_0}$, with $j_0\simeq 2-O(1/\sqrt\lambda)$.  With the elastic amplitude growing as a power, it follows that that   the dimensionless ratio $R_{el}$ also grows asymptotically  as $s^{j_0-1}$ thus violating the  constraint $R_{el}<1$ and unitarity correction must be taken into account.   Although the ``bare Pomeron'' approximation dominates in
the large $N_c$ expansion, it is clear that higher order summations are
necessary in order to restore unitarity.  To go beyond the
single-Pomeron approximation we must consider the high energy limit of
lower order terms in the $1/N_c$ expansion. 
  In flat space Veneziano
has shown that higher closed string loops for graviton scattering
eikonalize. Indeed in Refs. \cite{Brower:2007qh,Brower:2007xg} it was shown that the same sum leads to an eikonal
expansion that exponentiates for each string bit frozen  in impact parameter
during the collision.  To be  more explicit, the resulting  eikonal sum leads
to an impact representation for the 2-to-2 amplitude
\bea
A(s,x^\perp - x'^\perp)&=& - 2i s \int
 dz \ dz'\ P_{13}(z) P_{24}(z')  \left [
e^{i\chi(s,x^\perp - x'^\perp, z,z')} - 1\right]
\label{eq:adseik1}
\eea
  The eikonal $\chi$, as  a function of $x_\perp- x'_\perp$, $z,
z'$ and $ s$, can be determined by matching the first order term in $\chi$
to the
single-Pomeron contribution, (\ref{eq:adsPomeron}).
In impact space representation, and one finds
$
\chi(s,x_\perp- x'_\perp,z,z')= \frac{g^2_0 }{2 s} \; \widetilde {\cal  K}_P(s,x_\perp- x'_\perp, z,z')
$

This eikonal analysis can be extended directly to Higgs production. To  simplify the discussion, we shall adopt  a slightly formal treatment. Since  Higgs  is not part of  the  QCD dynamics, one can formally treat our eikonal as a functional of a
weakly coupled  external background Higgs field, $\phi_H(q^\pm,x^\perp_H,z_H)$, that is,  in (\ref{eq:adseik1}), we replace $A(s,x_\perp,x'_\perp)$ and    $\chi(s,x^\perp - x'^\perp, z,z')$ by $A(s,x^\perp, x'^\perp; \phi_H)$ and  $\chi(s,x^\perp - x'^\perp, z,z'; \phi_H)$,   with the understanding that they  reduce to $A(s,x_\perp,x'_\perp)$ and $\chi(s,x^\perp - x'^\perp, z,z')$ respectively in the limit $\phi_H \rightarrow 0$.  Since Higgs production is a small effect, by expanding to first order in the Higgs
background field, we find the leading order Higgs production amplitude, to all order in $\chi$, becomes
\bea
A_H(s_1, s_2,x^\perp -
x^\perp_H,x'^\perp -x^\perp_H, z_H) &=& 2 s \int
 dz \ dz'\ P_{13}(z) P_{24}(z')  \nn
 &\times&  \chi_H(s_1, s_2,x^\perp -
x^\perp_H,x'^\perp -x^\perp_H, z,z',z_H) \;    e^{i\chi(s,x^\perp - x'^\perp, z,z')}\nn
\eea
where $  \chi_H $ is  given by the Higgs production amplitude, (\ref{eq:adsDoublePomeronHiggs}),  due to double-Pomeron exchange in an impact representation~\footnote{To be more precise, up to a factor of $2s$,  $ \chi_H$ is given  by matching $A_H(s,x_\perp,x'_\perp, \phi_H)$  at
$|\chi | <<1 $ with the integrand  of (\ref{eq:adsDoublePomeronHiggs}) in an impact  representation.   We also note that here $s_1s_2/s\simeq m_H^2$.   The kinematics of transverse $AdS_3$ coordinates is represented schematically in Fig. \ref{fig:regge5}.}. The net effect of eikonal sum is to introduce a phase factor 
\begin{equation}
e^{i\chi(s,x_\perp- x'_\perp,z,z')}  = e^{i\; {\rm Re}\;\chi_R(s,x_\perp- x'_\perp,z,z')}e^{- {\rm Im} \; \chi(s,x_\perp- x'_\perp,z,z')}
\end{equation}
 into the production amplitude. Due to its absorptive part, ${\rm Im}\; \chi$, this eikonal factor provides a strong suppression for central Higgs production.  
 
 The effect of this suppression is often expressed in terms of a ``Survival Probability", $\langle S \rangle$.   In a momentum representation, the  cross section for Higgs production per unit of
rapidity in the central region is
\bea
\frac{d \sigma_H(s,y_H)}{dy_H} &=& \frac{ 1\;  }{ \pi^3 (16\pi)^2  s^2 } \int d^2q_{1\perp}  d^2q_{2\perp} |A_H(s,y_H,
q_{1\perp},q_{2\perp})|^2  
\label{eq:Hxsection}
\eea
where $y_H$ is the rapidity of the produced Higgs, $q_{1\perp} $ and $q_{2\perp} $ are transverse momenta of two outgoing fast leading particles  in the frame where the momenta of  incoming particles are longitudinal.  In (\ref{eq:Hxsection}), we have skipped writing integrals over the AdS radial directions.    ``Survival Probability" is conventionally defined by the ratio
\begin{equation}
\langle S^2 \rangle\equiv \frac{  \int d^2q_{1\perp}  d^2q_{2\perp} |A_H(s,y_H,
q_{1\perp},q_{2\perp})|^2  }{ \int d^2q_{1\perp}  d^2q_{2\perp} |A^{(0)}_H(s,y_H,
q_{1\perp},q_{2\perp})|^2}
\label{eq:survival}
\end{equation}
where $A^{(0)}_H$ is the corresponding amplitude before eikonal suppression, e.g., given by Eq. (\ref{eq:adsDoublePomeronHiggs}).  
For simplicity, we shall also focus on the  mid-rapidity production, i.e., $y_H\simeq 0$ in the overall CM frame. In this case, $\langle S^2 \rangle$ is a function of overall CM energy squared, $s$, or the equivalent total rapidity, $Y\simeq \log s$.   Evaluating the survival probability as given by (\ref{eq:survival}), though straight forward, is  often tedious. The structure for both the numerator and the denominator is the same. For numerator factor, one has
\bea
&&  \int
 dx_\perp dz \ d\bar z\ P_{13}(z) P_{13}(\bar z) \int d x'_\perp 
 dz' \ d\bar z'\ P_{24}(z) P_{24}(z') \int  e^{i\left(\chi (s,x_\perp-x'_\perp,z,z') - \chi ^*(s,x_\perp-x'_\perp,\bar z,\bar z') \right)}\nn
&&\chi_H(s,s_1, s_2,x^\perp -
x^\perp_H,x'^\perp -x^\perp_H, z,z') \chi_H^*(s,s_1, s_2,x^\perp -
x^\perp_H,x'^\perp -x^\perp_H,\bar z,\bar z')
\eea
where we have made use of that fact that  $z_H\simeq 1/m_H$. To obtain the denominator, one simply removes the phase factor, 
$
e^{i\left(\chi (s,x_\perp,x'_\perp,z,z') - \chi^*(s,x_\perp,x'_\perp,\bar z,\bar z') \right)}
$. It is now clear that it is this extra factor which controls the strength of suppression.

To gain a qualitative estimate, let us consider   the  local limit where $z\simeq \bar z\simeq z_0$ and  $z'\simeq \bar z'\simeq z'_0$, with $z_0\simeq z_0'\simeq 1/\Lambda_{QCD}$. In this limit,  one finds that this suppression factor reduces to
\begin{equation}
e^{- 2\; {\rm Im}\; \chi (s,x_\perp,x'_\perp,z_0,z'_0)}
\end{equation}
where $ {\rm Im}\; \chi>0$ by unitarity. If follows that, in a super-gravity limit of strong coupling where the eikonal is strictly real, there will be no suppression and the survival probability is 1. Conversely,  the fact that phenomenologically a small survival probability is required is another evidence that we need to work in an intermediate region where $1< j_0<2$.  In this more realistic limit, ${\rm Im}\; \chi$ is large and cannot be neglected.   In particular,  it follows that the dominant region for diffractive Higgs production in pp scattering comes from  the region where
 \begin{equation}
{\rm Im} \;\;\chi(s,x_\perp- x'_\perp,z,z') = O(1),
\end{equation}
with $z\simeq z' =O(1/\Lambda_{QCD})$. Note that this  is precisely the edge of the ``disk region"  for p-p scattering.   In order to carry out a quantitative analysis, it is imperative that we learn the property of $\chi(s,\vec b, z)$ for $|\vec b|$ large. From our experience with pp scattering, DIS at HERA, etc., we know that confinement will play a crucial role. In pp scattering, since $z\simeq z' =O(1/\Lambda_{QCD})$, we expect this condition is reached at relatively low energy, as is the case for total cross section. It therefore plays a dominant role in determining the magnitude of diffractive Higgs production at LHC.  We will not discuss this issue here further; more pertinent discussions on how to determine $\chi(s,x_\perp- x'_\perp,z,z')$ when confinement is important can be found in Ref. \cite{Brower:2010wf}.

We end by  re-iterating   the importance of factorization, Eq. (\ref{eq:adsDoublePomeronScheme}). Note that each separate factor in (\ref{eq:adsDoublePomeronScheme}), up to a proportionality constant, can also be related to the amplitude of proton scattering off a heavy "onium" state, with a mass $O(m_H)$, or, equivalently, that for  an off-shell Compton amplitude, e.g.,
\begin{equation}
\gamma^* \; (q_1)  + {\rm proton} \; (k_2)\rightarrow \gamma^* \; (q_3) + {\rm proton} \; (k_4)
\end{equation}
with $q_1^2 \simeq q_2^2 \simeq -m_H^2$. In this case, with $t$ small, the photon is highly virtual and the amplitude can be  thought of as a generalized DIS structure function, analogous to  the ``skewed  unintegrated'' parton distribution functions in a perturbative approach. (See, for instance, \cite{Ji:1996ek,Ji:1996nm,Ji:1998pc}.) This will allow one  to  relate the double diffractive  Higgs production amplitude  to other measurables in DIS, after implementing some plausible assumptions~\cite{Ellis:2008yp,Ross:2011zzb}.   It is also worth noting that  it should be possible to replace  the internal vertex $V_H$, or equivalently replace the top quark loop by
an operator for quark and anti-quark  production at the boundary of AdS space and thus extend the AdS  analysis to  Pomeron fusion into heavy quark di-jets production, providing a powerful experimental test and calibration  to the building blocks introduced for our diffractive Higgs production amplitude.  We are considering how to do this but leave this
extension as well as further phenomenological investigations to  future research.

\newpage

\vskip40pt

\noindent {\underline{Acknowledgments:}}
We are pleased to acknowledge useful conversations
with M. Block, M. S. Costa,  J. Ellis, E. Gotsman, H. Kowalski, E.  Levin, U. Maor,  D. A. Ross, M. Strassler,  and C. Vergu.  The
work of R. C. B.  was supported by the Department of Energy under
contract~DE-FG02-91ER40676, and that of  C.-IT.  by the
Department of Energy under contract~DE-FG02-91ER40688, Task-A. R.B. and C.-IT. would like to thank the Aspen Center
for Physics for its hospitality during the early phase  of this work. Centro de F\'isica do Porto is partially funded by FCT and the work of M.D. is partially supported by grants PTDC/FIS/099293/2008 and CERN/FP/116358/2010 and by the FCT/Marie Curie Welcome II project.

\newpage
\bibliographystyle{utphys}
\bibliography{PomeronHiggs}


\newpage
\def\theequation{A.\arabic{equation}}
\setcounter{equation}{0}

\appendix

\newpage
\section{AdS QCD and Conformal Symmetry Breaking}
\label{sec:Confinement}

In this appendix, we first present a model with scale invariance breaking due to confinement deformation, leading to  non-vanishing expectation values  for $\langle
F^2\rangle$, etc.  Under such a condition,   there will be a Witten  graviton-graviton-dilaton vertex  in the bulk as well as the existence of tensor glueballs. We next turn to a more amenable model, e.g., the hard-wall model, where, due to confinement, the existence of a tensor glueball can be more easily studied. This approach can also be used to provide a more explicit representation for our Pomeron kernel, ${\cal K}_P$. Lastly, we discuss consequence of asymptotic freedom, and point out how a non-vanishing beta-function can be used to relate gluon condensate  to the non-vanishing  trace for the energy-momentum tensor. This observation has been used to provide an estimate on the magnitude for the  central Higgs production vertex $V_H$ in Sec.~\ref{sec:strategy}.

\paragraph{Symmetry Breaking and AdS/CFT Correspondence:} 
Symmetry breaking effect  has been studied in AdS/CFT correspondence mostly using a near-boundary analysis~\cite{Witten:1998qj,Gubser:1998bc,Balasubramanian:1998sn,Banks:1998dd,Klebanov:1999tb, Polchinski:2000uf,Skenderis:2002wp}.  In a standard AdS/CFT treatment for $d=4$, ${\cal N}=4$ Yang-Mills theory, each local operator ${\cal O}_i$ of dimension $\Delta_i$ in the CFT  corresponds to two possible solutions of the linearized field equations  for an associated bulk field $\Phi_i$ near the $AdS$ boundary~\cite{Witten:1998qj,Gubser:1998bc}, a nonnormalizable solution which scales with the AdS radius as $r^{\Delta_i-4}$ with $r$ the $AdS$ radius, and a normalizable solution which scales as $r^{-\Delta_i}$.   As explained carefully in \cite{Polchinski:2000uf}, when a CFT is perturbed by a symmetry breaking local operator $  a_i{\cal O}_i$, a non-vanishing vev $\langle{\cal O}_i \rangle$ corresponds to a supergravity solution which behaves at large $r$ as
\begin{equation}
 a_i r^{\Delta_i-4} +\cdots + b_i \; r^{-\Delta_i}+\cdots
\end{equation}
where $\langle{\cal O}_i\rangle= b_i$.  For our present purpose, it is more profitable to address scale invariance braking in terms of Witten diagrams in the bulk. 

In a truly realistic treatment of  QCD, one expects scale invariance breaking due to running of QCD coupling, with the trace of the stress-energy  tensor  related to  $F^2$  by the QCD beta-function, e.g., $\langle T^\mu_\mu\rangle =\beta \; \langle F^2\rangle $. 
In an approximate treatment where  conformal invariance is maintained, on the other hand, it follows that $T_{\mu\nu}$ remains traceless, with vanishing $\langle F^2\rangle=0$, etc.  More generally, we will be interested in n-point correlators involving $F^2$ and $T_{\mu\nu}$, which   can  be evaluated at strong coupling through the use of Witten diagrams.   Under such a scenario, as we shall explain below by an explicit model for dilaton-gravity,   the relevant Graviton-Higgs-Graviton coupling in the bulk would vanish. That is, there would be no double diffractive Higgs production at strong coupling! Fortunately,  in a framework with proper confinement deformation, scale invariance is broken with non-vanishing vev for $F^2$, and the Graviton-Higgs-Graviton coupling is non-zero. Furthermore, when the running coupling is properly taken into account, through the conformal anomaly, we are able to fix the normalization of double diffractive Higgs production. 

\paragraph{Central Vertex and Scale Invariance Breaking:}

The importance of scale invariance breaking for the QCD dynamics has been  emphasized in \cite{Hashimoto:1998if}. It has also been discussed by many  in connection with the spectrum of  light scalar glueballs. It is possible to adopt a holographic approach where one couples 5d gravity to a dilaton with $AdS_5$ geometry in the UV, Eq. (\ref{eq:nearAdS}) while taking on a non-trivial background. For instance, this can be achieved by considering a toy model with action~\cite{Csaki:2006ji}
\begin{equation}
S=M^2_P \int d^5 x \sqrt g \Big( -{\cal R} - V(\phi) + \frac{1}{2} g^{MN} \dd_M\phi \dd_N \phi - \lambda(\phi) T(z) \Big)
\end{equation}
where $\phi$ is the dilaton and $V(\phi)$ is a dilatonic potential and we have added an external ``brane'' source. This system was analyzed in some detail in DeWolfe, Freedman, Gubser and Karch~\cite{DeWolfe:1999cp} by mapping the problem into the solution of a super potential.  
A first question is, without introducing an external brane source, but with a potential $V$ appropriately chosen, is it possible to simulate running coupling, scale invariance breaking, etc?  This leads to a non-vanishing gluon condensate, and, in such a setting, a non-vanishing graviton-graviton-dilaton vertex can be obtained.  Since double diffractive Higgs production proceeds via scalar glueball production, as discussed earlier, this in turn leads to a non-vanishing Pomeron-Pomeron-Higgs vertex, $V_H$.

Consider first an example   leading  to  a ``near-$AdS$'' geometry, (\ref{eq:nearAdS}), by having   a dilaton background, $\phi_{cl}(z)$, which can be non-trivial.  Expanding the action about this background, $G_{mn}=g_{mn} + h_{mn}$ and $\phi=\phi_{cl}+ \varphi$, and consider traceless-transverse $h_{mn}$ appropriate for Pomeron/Graviton fluctuations, terms linear in $h$ and $\varphi$ vanish, quadratic terms serve to determine the glueball spectrum, and
higher order terms  correspond to couplings in a Witten diagrammatic treatment. We are therefore interested in a term linear in $\varphi$ and quadratic in $h$. Expanding the action we obtain terms of this order
\begin{equation}
S_{int} = \frac{M^2_P}{4 } \int dz d^4x \sqrt{-g}
h^{nm}h_{mn} [V'(\phi_{cl}) \varphi - g^{zz} \partial_z
\phi_{cl}(z) \partial_z \varphi]  \;.
\end{equation}
where $\phi_{cl}(z)$ is the classical background. Note in the pure $AdS^5$ conformal  background, the potential is a constant, $V = -12/R^2$, i.e., it simply provides the cosmological constant, and $\phi_{cl}(z) = \mbox{const}$. It follows both terms above  are
zero.  That is, the graviton-graviton-dilaton coupling vanishes identically in a conformal limit. 

Conversely, if the background is non-trivial, and breaks scale invariance, a more involved potential $V(\phi)$ is required.
 This in turn leads to a non-zero graviton-graviton-dilaton coupling. This can best be illustrated for example by  the special ``subcritical asymptotically free'' solution,  discussed in Csaki et al.~\cite{Csaki:2006ji}, where
\begin{equation}
V(\phi) = - \frac{6}{R^2} e^{\textstyle \sqrt{2/3} \phi} -
\frac{12}{R^2}\quad , \quad
\phi_{cl} = - \sqrt{3/2} \; \log \log (z_0/z)
\end{equation}
We note that, for this case, 
\begin{equation}
e^{\textstyle \sqrt{2/3} \phi_{cl}} = 1/\log (z_0/z)  
\end{equation}
 As an illustration of the effects of running coupling in the UV, this model
gives
\begin{equation}
S_{int} =  - \frac{M^2_P}{4} \int dz d^4x \sqrt{-g} \frac{\sqrt{6}}{R^2\log (z_0/z)}
h^{nm}h_{mn} [  \phi + \frac{z}{ 4}
\partial_z \phi]
\label{eq:WittenVertex}
\end{equation}
with a non-trivial graviton-graviton-dilaton coupling in the bulk. 

Although the model discussed above looks attractive, it nevertheless cannot serve as a realistic model for QCD as we will indicate shortly. The precise choice of the best background is therefore  left to future phenomenology. Here it is sufficient to assume that
there is confining background so that there is a stable (at large $N$) tensor glueball on the leading trajectory at $J = 2$  and
a non-zero Pomeron-Pomeron-dilaton vertex. As we shall see shortly, the existence of a tensor glueball plays a crucial role in our treatment, we turn next to a more detailed analysis using a hard-wall background which is also more amenable to a $J$-plane analysis, necessary for discussing our Pomeron kernel. 

\paragraph{Tensor Glueballs and Confinement Deformation:}

Let us begin by first considering  confinement deformations keeping the ultraviolet region conformal. If confinement sets in at a scale $\Lambda$ in the gauge theory, this leads to a change in the metric away from $AdS_5$ in the region near $z=R^2/r \sim 1/\Lambda=z_0$. One can think of the space being ``cut-off" or ``rounded off", in some natural way at $z=z_0$, leading to a ``wave-guide" effect. This in general leads to a theory with a discrete hadron spectrum, with mass splitting of the order $\Lambda$ among hadrons of spin $\leq 2$.  As discussed in \cite{Brower:2006ea}, the differential operator determining the $J$-plane spectrum remains approximately unaffected for $-t>> \Lambda^2$, while the effect of confinement becomes important as $t\rightarrow 0^{-}$, and for any $t>0$.   To  gain a qualitative understanding, it is instructive to treat below the ``hard-wall" model, so that Eq. (\ref{eq:tensor}) remains valid, with $z$ cutoff in the range $[0,z_0]$.  While this model is not a fully consistent theory, it does capture key features of confining theories with string theoretic dual descriptions. 

To set the proper stage, let us first review the situation in the conformal limit. Recall that, at   finite $\lambda$, it has been shown in Ref. \cite{Brower:2006ea} that, due to curvature of AdS, the effective spin of a graviton exchange is lowered from 2 to $j_0=2-2/\sqrt \lambda$. As such it is necessary to adopt a $J$-plane formalism where the Pomeron kernel $\widetilde {\cal K}_P$  is given by an inverse Mellin transform, 
\begin{equation}
\widetilde{\cal K}_P(s,t,z,z') =
-  \int_{-i\infty}^{i\infty} \frac{dj}{2\pi i} (
\alpha'  \widehat s)^{j} \frac{1 + e^{-i \pi j}}{\sin\pi j}   \widetilde G_j(t,z,z') \; .
\end{equation}
with $\widehat s= zz's/R^2$. Note that  the $J=2$ contribution is
\be
(2/\pi) (
\alpha'  \widehat s)^{2}  \widetilde G_2(t,z,z') \; .
\label{eq:graviton2}
\ee
This  is simply the graviton Kernel, (\ref{eq:graviton}), up to a constant factor, which can be absorbed into the coupling. 
When conformal invariance is maintained, this $J$-dependent propagator  $\widetilde G_j(t,z,z')$ satisfies the standard $AdS_5$ differential equation
\bea
\left( -z\partial_z z\partial_z + (2\sqrt{\lambda})(j-j_0) - z^2 t\right) \widetilde  G_j(z,z';t) &=& z \; \delta(z-z') 
\eea
with  $j_0=2-2/\sqrt \lambda$.    
This equation can also be expressed in a standard Schrodinger form. It can be solved either by a spectral resolution in $t$ or in $j$. Holding $j>j_0$ first and real,  the spectrum in $t$ can be seen to be continuous, along its positive real axis, leading to 
\bea
 \widetilde G_j(z,z';t)
&= & \int_0^\infty kdk \frac{ J_{(\Delta(j)-2)}(kz)J_{(\Delta(j)-2)}(kz')}{k^2-t}\;. \label{eq:spectrumq3}
\eea
where $\Delta(j)=2+ \sqrt {2\sqrt{\lambda} (j-j_0)}$. 
If, on the other hand,  an IR hard-wall cutoff is introduced, the spectrum in $t$  becomes discrete.   The propagator is now given by a discrete sum,
\bea
\widetilde G_j(z,z';t)
&=&  \sum_n \frac{ \widetilde \phi_n(z,j)\widetilde \phi_n(z',j) }{m^2_n(j) -t} \;.
\eea
The wave-functions $\widetilde \phi_n(z,j)$ can again be expressed in terms of Bessel functions, 
with $m_n(j)$ fixed by a Neumann condition at $z_{0}$, $[z^j\widetilde \phi_n(z,j)]'\big|_{z=z_0}=0 $.   This discrete structure is what is to be expected when confinement deformation is introduced.

Alternatively, one  finds that
\begin{equation}
 \widetilde G_j(z,z';t) =  \int_{-\infty}^\infty  \frac{d\nu}{\pi^2} (\nu \sinh\pi \nu) \frac{K_{i\nu}(q z) K_{i \nu}(q z')}{ 2 \sqrt{\lambda}(j - j_0) + \nu^2}   \label{eq:adsPompropagatorJ}
\end{equation}
where $t=-q^2<0$. (\ref{eq:adsPompropagatorJ}) provides a different representation for the Pomeron propagator in the conformal limit. One observes more directly the presence of the branch cut at $j=j_0$, which corresponds to the minimum of the denominator in (\ref{eq:adsPompropagatorJ}).  From (\ref{eq:spectrumq3}), the presence of this cut has to be inferred from the dependence  through $\Delta(j)$.  
Equivalently, one can work directly with 
a spectrum analysis in the $J$-plane, as carried out in  Ref. \cite{Brower:2007xg}. 
The analysis can best be thought of as working with the boost operator, $\hat H = M_{+-}$, which is  conjugate to $J$.

 It is also just as easy to treat the problem more generally,  e.g., with $ e^{\textstyle  -2 A(z)}$ replacing $(R/z)^2$ as a confining warping factor.
The Green's function is the spectral representation for the
operator, $(j - \hat H)^{-1}$,
\begin{equation}
G_j(q_\perp, z,z') = e^{\textstyle  -j A(z)}
 \sum_n\frac{\psi_n(q_\perp, z) \psi_n(q_\perp, z')}{j - \lambda_n(q)} e^{\textstyle  -j A(z')}
\label{eq:spectrumqIRGeneral}
\end{equation}
where traditional one regards $\omega_n = \lambda_n -1$ as eigen-energies
of boost operator~\footnote{Actually confinement alone has
both a discrete and continuum spectrum, which we do not exhibit explicitly here. }.

\begin{figure}[bthp]
\begin{center}
\includegraphics[width=0.60\textwidth]{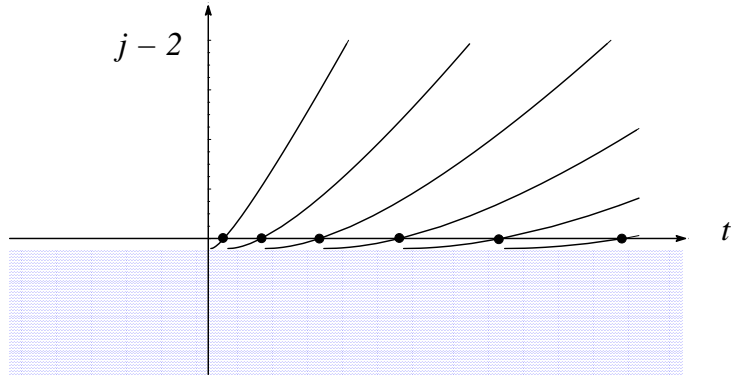}
\end{center}
\caption{The analytic behavior of  Regge trajectories in the hard-wall model,
  showing the location of the bound-state poles at $j = 2$ and the
  $t$-independent continuum cut (shaded) 
  at $j=j_0=2-2/\sqrt{\lambda}$ into which the Regge
  trajectories disappear.  The lowest Regge trajectory intersects the
  cut at a small positive value of $t$.  At sufficiently large $t$
  each trajectory attains a fixed slope, corresponding to the tension
  of the model's confining flux tubes.}
\label{fig:hardwall}
\end{figure}

Further discussion on these construct can be found in Ref. \cite{Brower:2006ea}.  
 Here, we simply display in Fig.~\ref{fig:hardwall}, the structure of these Regge singularity  for  $J$ and $t$ real. In particular, we emphasize that, when the trajectory crosses $j=2$, it corresponds to a physical tensor glueball. This important feature we shall make use of in the next section.

\begin{figure}[bthp]
\begin{center}
\includegraphics[width = 0.60\textwidth]{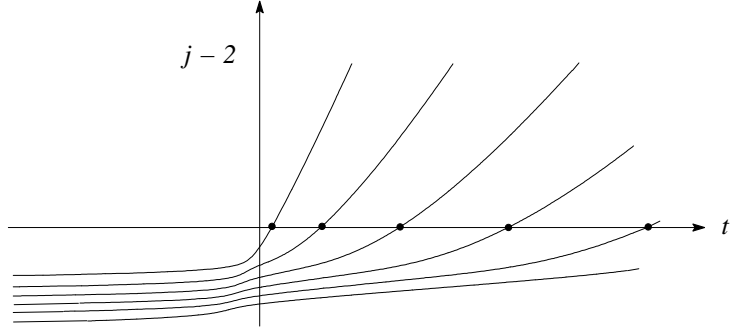}
\end{center}
\caption{Discrete Regge Spectrum at negative $t$ due to running coupling}
\label{fig:evrunning}
\end{figure}

It is possible to adopt a more general background which mimics the features of running coupling. This has also been done in \cite{Brower:2006ea}. The result is a discrete spectrum in the Regge plane in agreement with expectation based on 
asymptotic freedom for the BRST equation,  Fig.~\ref{fig:evrunning}.

\paragraph{ Modeling Asymptotic Freedom and Broken Scale Invariance:}

In a realistic holographic treatment of QCD, asymptotic freedom must be handled consistently. Although there have been several models
to implement  some features of running coupling ~\cite{Brower:2006ea,Csaki:2006ji} , they nevertheless do not prove to be entirely  inadequate. For instance, they
fail to give   the correct running coupling dependence as measured by the potential between static quarks  at a small separation $L$
\begin{equation}
V_{\bar QQ}(L) \simeq  - \alpha(L)/L  \quad, \quad  \alpha(L) = \frac{g^2 N_c}{- \log(\mu L )}
\end{equation}
when computed using  the Nambu action  in these deformed backgrounds. 
We have checked that the  background of \cite{Csaki:2006ji}  with running coupling does not support this interpretation.  An analogous
suggestion  in ~\cite{Brower:2006ea}, discussed briefly above,    does lead
closer to the desired answer. However it is not
a solution to pure dilaton gravity. It requires a negative tension brane in the UV to support it
or some other source of energy. Perhaps it is not surprising that to wed the UV behavior
to a smooth effective background at strong coupling should be difficult. A real
QCD dual theory would have to describe hard scattering and gluon jets at high energy
not just a running coupling. Most likely this at the very least implies highly curved background far from these phenomenological attempts.

Clearly the
details of such a construct  go beyond the scope of the current discussion. It suffices to emphasize that $V_H\neq 0$ is a general consequence of scale anomaly. As pointed out earlier,  in a naive application of the  hard-wall model, such vertex is indeed absent in the bulk. Therefore, our discussion of Higgs production in AdS/CFT should be understood in a more general setting with a proper confinement deformation of the AdS geometry.

\end{document}